\documentclass[%
 reprint,
 amsmath,amssymb,
 aps,
]{revtex4-2}

\usepackage{graphicx}
\usepackage{dcolumn}
\usepackage{bm}

\usepackage{todonotes}

\begin{document}

\preprint{APS/123-QED}

\title{Active wave-particle clusters}

\author{Rahil N. Valani$^{1}$}\email{rahil.valani@physics.ox.ac.uk}
\author{David M. Paganin$^{2}$}\email{david.paganin@monash.edu}
\affiliation{$^1$Rudolf Peierls Centre for Theoretical Physics, Parks Road,
University of Oxford, OX1 3PU, United Kingdom}
\affiliation{$^2$School of Physics and Astronomy, Monash University, Victoria 3800, Australia}

\date{\today}

\begin{abstract}
Active particles are non-equilibrium entities that uptake energy and convert it into self-propulsion. A dynamically rich class of inertial active particles having features of wave-particle coupling and wave memory are walking/superwalking droplets. Such classical, active wave-particle entities (WPEs) have previously been shown to exhibit hydrodynamic analogs of many single-particle quantum systems. Inspired by the rich dynamics of strongly interacting superwalking droplets in experiments, we numerically investigate the dynamics of WPE clusters using a stroboscopic model. We find that several interacting WPEs self-organize into a stable bound cluster, reminiscent of an atomic nucleus. This active cluster exhibits a rich spectrum of collective excitations, including shape oscillations and chiral rotating modes, akin to vibrational and rotational modes of nuclear excitations, as the spatial extent of the waves and their temporal decay rate (memory) are varied. Dynamically distinct excitation modes create a common time-averaged collective wave field potential, bearing qualitative similarities with the nuclear shell model and the bag model of hadrons. For high memory and rapid spatial decay of waves, the active cluster becomes unstable and disintegrates; however, within a narrow regime of the parameter space, the cluster ejects single particles whose decay statistics follow exponential laws, reminiscent of radioactive nuclear decay. Our study uncovers a rich spectrum of dynamical behaviors in clusters of active particles, opening new avenues for exploring hydrodynamic quantum analogs in active matter systems. 

\end{abstract}

\maketitle


\section{Introduction}

The evolution of systems exhibiting memory effects is often governed by integrodifferential equations, which abound in the physics of coupled particle--field systems. For example, in classical electrodynamics, the Lorentz force acting on each member of a point-charge system leads to an integrodifferential equation for the motion of each particle~\cite{StrazhevShkolnikov1989}. As another example, the Dirac equation assumes an integrodifferential form when a single-electron Dirac field is coupled to an electromagnetic field created by the past history of the system~\cite{DeWitt1962}. For a classical particle coupled to the field it radiates, the particle interacts locally (i.e., via a finite-order differential equation) with the field created by its past history (as described by an integral superposing Green functions associated with the past states of the system). Of course, when the effects of radiation reaction are weak, one can work with finite-order differential equations by introducing the simplifying approximation that the particle is the ``source'' which radiates an associated field described via electromagnetic potentials \cite{WaldElectrodynamicsBook,JacksonBook,Dirac1938}. However, when backreaction is strong---i.e., when the particle and its associated field exhibit significant mutual influence upon one another---it is often natural to retain the full complexity of the governing integrodifferential equation. Alternatively, reduced integrodifferential equations may be employed, with examples including the Hartree--Fock equation of elementary quantum mechanics \cite{MerzbacherQMbook} and the ``IDEA'' formalism in nuclear physics \cite{IDEA00,IDEA0,IDEA1,IDEA2,IDEA3,IDEA4}.

\begin{figure}
\centering
\includegraphics[width=0.9\columnwidth]{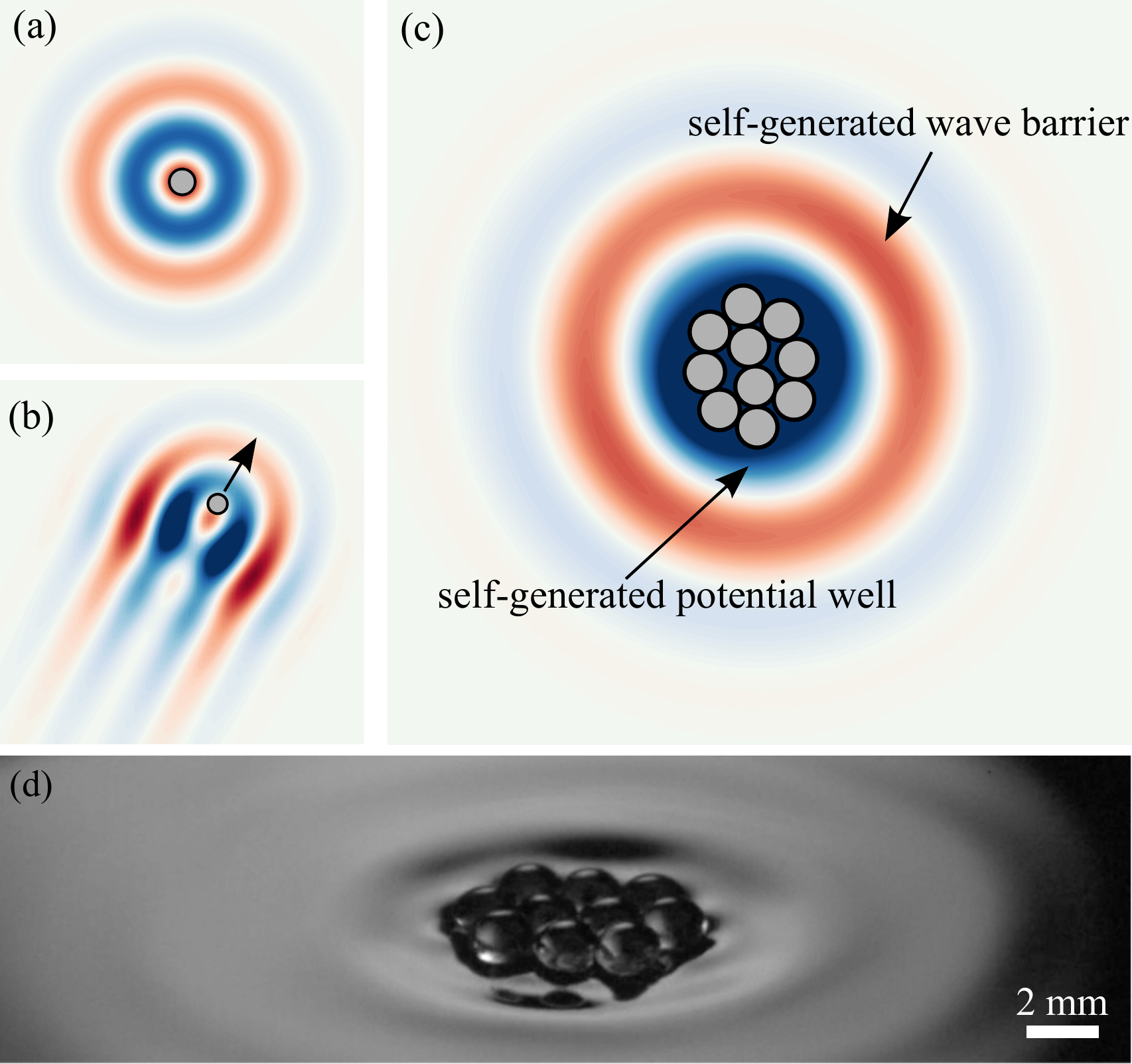}
\caption{Active wave-particle cluster. A finite-size particle (gray) generates (a) an axisymmetric wave of form $W(|\mathbf{x}|)=\cos(|\mathbf{x}|)\exp[{-(|\mathbf{x}|/L)^2}]$ (peaks in red and troughs in blue) at each instant of time. (b) The superposition of all the individual waves generated by the particle results in an overall wave field that propels an isolated particle with constant velocity (black arrow), making it active. (c) Simulated snapshot in time of a collection of such active WPEs that self-organize into a ``nucleus" structure, where the particles stay bound by their self-generated collective wave field comprising a potential well walled by a wave barrier. (d) Experimental image of a self-organized cluster of superwalking droplets~\citep{superwalker}.}
\label{Fig: schematic}
\end{figure}

A particular class of matter, known as active matter, has gained considerable physics interest in recent years~\citep{doi:10.1146/annurev-conmatphys-070909-104101}. Here the individual units are non-equilibrium active particles that consume energy and convert it into persistent motion. Examples includes animate matter from motile microscale bacteria to animals and birds, as well as inanimate matter such as active colloids and microrobots~\citep{Pismen2021_Ch3,Pismen2021_Ch4}. An inanimate hydrodynamic active system that exhibits memory effects and can also be described by integrodifferential equations is that of walking/superwalking droplets~\citep{Couder2005WalkingDroplets,protiere_boudaoud_couder_2006,superwalker}. In this system, a droplet of oil self-propels horizontally while periodically bouncing on a vertically vibrated bath of the same oil. Each bounce of the droplet excites localized damped standing waves which the droplet interacts with on subsequent bounces to propel itself. Such walking droplets are active particles because they locally consume energy from global vertical bath vibrations and convert it into persistent horizontal self-propulsion. This active system is dynamically rich, with two key features: (i) the droplet and its self-generated wave coexist as a \emph{wave-particle entity} (WPE), wherein the droplet creates localized waves which in turn guide the droplet motion, and (ii) the system is non-Markovian and has \emph{memory} since the waves generated by the droplet can decay very slowly in time and hence the droplet motion is affected by the history of waves along the droplet's trajectory. Such WPEs exhibit hydrodynamic quantum analogs~\citep{Bush2015,Bush_2020,Bush2024,10.1063/5.0210055} and some examples include quantization of orbits~\citep{Fort17515,harris_bush_2014,Oza2014,Perrard2014b,Perrard2014a,labousse2016}, tunneling across submerged barriers~\citep{Eddi2009,tunnelingnachbin,tunneling2020}, wave-like statistics in confined geometries~\citep{PhysRevE.88.011001,Giletconfined2016,Saenz2017,Cristea,durey_milewski_wang_2020}, Friedel oscillations~\citep{Friedal} and hydrodynamic superradiance~\citep{Superradiance2021,PhysRevLett.130.064002}. 
 
Most studies on hydrodynamic quantum analogs have focused on single WPEs, with limited work on strongly-interacting systems \cite{Helias2023,Simula2023-lh,Couchman_Bush_2020,Protiere_2005,Eddi_2011}. By contrast, it is well known that active matter systems exhibit a diverse range of collective behaviors when many active particles interact~\citep{doi:10.1146/annurev-conmatphys-070909-104101}. In particular, they exhibit clustering, from few-particle assemblies to mesoscopic aggregates \cite{RevModPhys.85.1143,Bechinger2016,doi:10.1146/annurev-conmatphys-031214-014710}. Minimal clusters such as Quincke roller clusters \cite{Mauleon-Amieva2021} and phoretic Janus colloid rotors \cite{Theurkauff2012} reveal orbiting, translating, and rotating states. 

In this paper, we investigate the rich dynamics of strongly-interacting active WPE clusters, and in certain regimes we establish some similarities with nuclear physics. Our paper is organized as follows. In Sec.~\ref{Sec: Model} we present the stroboscopic theoretical model for many interacting WPEs, followed by a description of the formation of an active WPE cluster in Sec.~\ref{Sec: Cluster}. In Sec.~\ref{Sec: PS space} we explore in detail the different dynamical behaviors observed for the WPE clusters in different regions of their associated parameter space, while making connections with nuclear physics and active matter. We conclude with Sec.~\ref{Sec: DC}.
 

\section{Active wave-particle model}\label{Sec: Model} 

We consider $N$ identical droplets (particles) bouncing in phase periodically on a vertically vibrating bath of the same liquid while moving horizontally in two dimensions $\mathbf{x}=(x,y)$. By averaging over the fast time scale of vertical bouncing, \citet{Oza2013} developed a theoretical stroboscopic model describing the horizontal motion of such a wave-particle entity (WPE) as an integrodifferential equation. The equation of motion for the two-dimensional horizontal dynamics of a single WPE is
\begin{widetext}
\begin{align}\label{eq: dimensional_2D}
    m \ddot{\mathbf{x}}_p + D \dot{\mathbf{x}}_p = 
    -\frac{mg A_0 k_F}{T_F} \int_{-\infty}^{t} W'\left( k_F |\mathbf{x}_p(t)-\mathbf{x}_p(s)| \right)\frac{\mathbf{x}_p(t)-\mathbf{x}_p(s)}{|\mathbf{x}_p(t)-\mathbf{x}_p(s)|}\,\text{e}^{-\frac{(t-s)}{T_F \text{Me}}}\,\text{d}s,
\end{align}
\end{widetext}
where $m$ is the droplet mass, $D$ is an effective drag coefficient, $g$ is gravitational acceleration, $A_0$ is the amplitude of the droplet-generated surface waves, $k_F = 2 \pi / \lambda_F$ is the Faraday wavenumber corresponding to the Faraday wavelength $\lambda_F$, and a dash denotes differentiation with respect to the argument of a function. $T_F$ is the Faraday period (i.e., the period of the droplet-generated standing waves) and $\text{Me}$ is the memory parameter controlling the decay rate of droplet-generated surface waves. We refer the reader to \citet{Oza2013} for more details and explicit expressions for these parameters. The function $W(|\mathbf{x}|)$ represents the axisymmetric wave field generated by each particle at each instant. The left-hand side of Eq.~\eqref{eq: dimensional_2D} is composed of an inertial term $m \ddot{\mathbf{x}}_{p}$ and an effective drag force $D \dot{\mathbf{x}}_{p}$. The term on the right-hand side captures the forcing on the particle from its self-generated wave field. This force is proportional to the gradient of the self-generated wave field and it is calculated through integration of the individual wave forms $W(\mathbf{|x|})$ that are continuously generated by the particle along its trajectory and decay exponentially in time.

For many interacting WPEs, the horizontal dynamics of the $i$th particle is influenced by both its self-generated wave field and the wave field of other particles. Moreover, we consider a short-range repulsive force $F_{ij}^{rep}$ as an excluded volume interaction to account for the finite particle size. Thus, the equation of motion for many interacting WPEs is
\begin{align}\label{eq: dimensional_multi}
m \ddot{\mathbf{x}}_{pi} + D \dot{\mathbf{x}}_{pi} = F_{ii}^{int} + \sum_{\substack{j=1 \\ j\neq i}}^{N} (F_{ij}^{int} + F_{ij}^{rep}),
\end{align}
with
\begin{widetext}
\begin{align*}
F_{ij}^{int} &= - \frac{mg A_0 k_F}{T_F} \int_{-\infty}^{t} W'\left( k_F |\mathbf{x}_{pi}(t)-\mathbf{x}_{pj}(s)| \right)\frac{\mathbf{x}_{pi}(t)-\mathbf{x}_{pj}(s)}{|\mathbf{x}_{pi}(t)-\mathbf{x}_{pj}(s)|}\,\text{e}^{-\frac{(t-s)}{T_F \text{Me}}}\,\text{d}s, \\
F_{ij}^{rep} &= \bar{K} \frac{\mathbf{x}_{pi}(t)-\mathbf{x}_{pj}(t)}{|\mathbf{x}_{pi}(t)-\mathbf{x}_{pj}(t)|} \left(\bar{a} - |\mathbf{x}_{pi}(t)-\mathbf{x}_{pj}(t)|\right),
\end{align*}
\end{widetext}
when $|\mathbf{x}_{pi}-\mathbf{x}_{pj}|<\bar{a}$ and $F_{ij}^{rep}=0$ otherwise, with $\bar{a}$ as the particle diameter. The first term on the right-hand side, $F_{ii}^{int}$, captures the forcing on the particle from its self-generated wave field. The second term, $F_{ij}^{int}$, represents the force on the $i$th particle due to the wave field of the $j$th particle, while the last term, $F_{ij}^{rep}$, is a short-range repulsive spring force between particles $i$ and $j$, with spring constant $\bar{K}$, capturing the finite size of the particle. We choose the wave form 
\begin{equation*}
W(|\mathbf{x}|) = \cos(k_F|\mathbf{x}|)\exp[-(|\mathbf{x}|/l)^2],
\end{equation*}
where $l$ is the length scale of spatial decay (see Fig.~\ref{Fig: schematic}(a)). The state-of-the-art experimentally measured wavefield~\citep{Damiano2016} has features of both (i) spatial oscillation, given by the $\text{J}_0(\cdot)$ Bessel function, and (ii) spatial decay given by the $\text{K}_1(\cdot)$ Bessel function. Our chosen wave form, constructed from the more elementary sinusoidal and Gaussian functions, preserves the essential qualitative features of the experimental wavefield, where the wavelength of the sinusoidal spatial oscillations is given by $\lambda_F$ and the length scale of the spatial Gaussian decay by $l$. Moreover, this choice is consistent with our previous work on one-dimensional WPEs, where the corresponding dynamical regimes and linear stability calculations are well characterized~\citep{Valaniunsteady2021}. This approach aligns with the broader goal of the present study, which is to investigate the complex dynamical behaviors emerging in strongly interacting WPE-cluster systems across parameter space, rather than to provide quantitative predictions directly comparable to experimental observations in one particular system.

We non-dimensionalize the equation of motion using the characteristic length scale and time scale:
\begin{equation*}
\mathbf{x}' = k_F \mathbf{x}, \quad t' = \frac{D}{m} t.
\end{equation*}
Dropping primes on the dimensionless variables, the dimensionless equations of motion read
\begin{align}\label{eq: nondim_multi}
\ddot{\mathbf{x}}_{pi} + \dot{\mathbf{x}}_{pi} = F_{ii}^{int} + \sum_{\substack{j=1 \\ j\neq i}}^{N} (F_{ij}^{int} + F_{ij}^{rep}),
\end{align}
with
\begin{align*}
&F_{ij}^{int} = \\
&- \! R \! \int_{-\infty}^{t} \!\!\!\!\!\! W'\left( |\mathbf{x}_{pi}(t)-\mathbf{x}_{pj}(s)| \right)\frac{\mathbf{x}_{pi}(t)-\mathbf{x}_{pj}(s)}{|\mathbf{x}_{pi}(t)-\mathbf{x}_{pj}(s)|}\,\text{e}^{-\frac{(t-s)}{\tau}}\,\text{d}s,
\end{align*}
where the dimensionless wave amplitude and memory time are
\begin{equation*}
R = \frac{m^3 g A_0 k_F^2}{D^3 T_F}, \quad \tau = \frac{D T_F \text{Me}}{m}.
\end{equation*}
The dimensionless short-range repulsive force $F_{ij}^{rep}$ now has dimensionless particle diameter $a=k_F \bar{a}$ and the dimensionless spring constant is $K=\bar{K} m/D^2$. The dimensionless waveform reads
\begin{equation*}
W(|\mathbf{x}|) = \cos(|\mathbf{x}|) \exp[-(|\mathbf{x}|/L)^2],
\end{equation*}
where $L=k_F l$ is the dimensionless spatial decay length of the waves. In the main text of this paper, we explore in detail the dynamics of $N=10$ interacting WPEs and consider two different particle diameters based on (i) the typical particle diameter of a single-frequency driven walker~\citep{Molacek2013DropsTheory}, namely $\bar{a}=0.8$\,mm, and (ii) a two-frequency driven superwalker~\citep{superwalker,superwalkernumerical}, with $\bar{a}=1.3$\,mm. Considering a typical value of the Faraday wavelength to be $\lambda_F\approx 5$\,mm for walking-droplet experiments~\citep{Molacek2013DropsTheory}, this corresponds to $a=1.01$ and $R=0.77$ for a walker, and $a=1.63$ and $R=1.72$ for a superwalker. We refer the reader to Appendix \ref{App 1} for details of the numerical implementation.  Note, also, that the effects of varying $N$ are briefly considered in Appendix \ref{sec: diff para}.

\begin{figure*}
\centering
\includegraphics[width=1.8\columnwidth]{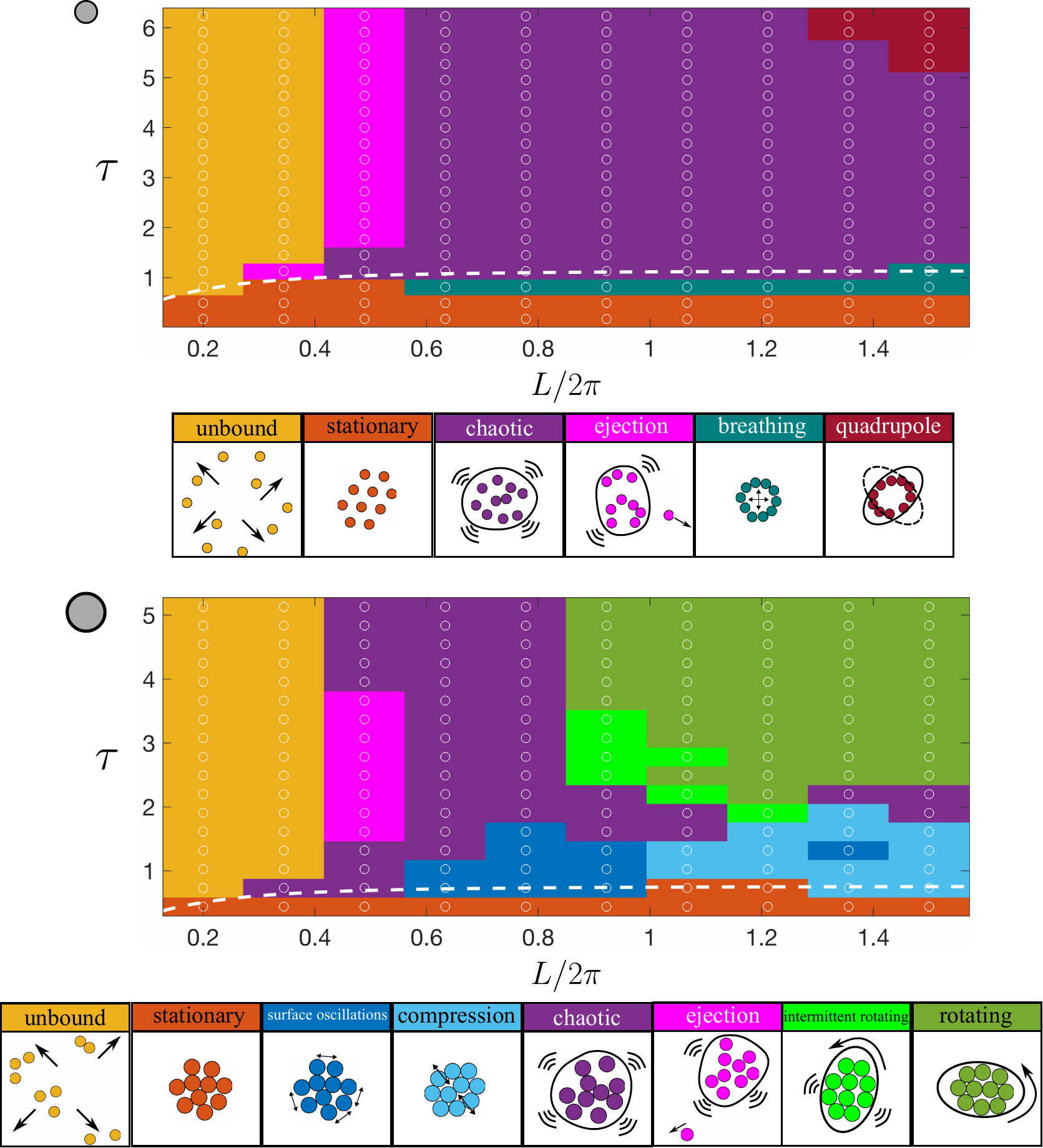}
\caption{Dynamical regimes for active wave-particle clusters having $N=10$ particles. Different dynamical behaviors for the active wave-particle cluster in the parameter space formed by the dimensionless spatial decay length scaled with the Faraday wavelength, $L/2\pi$, and the dimensionless memory parameter, $\tau$, for small particles of dimensionless diameter $a=1.01$ with $R=0.77$ (upper part of the figure) and larger particles of $a=1.63$ with $R=1.72$ (lower part of the figure). The white dashed curve in both parts denotes the critical memory curve $\tau_c=1/\sqrt{{R\left(1+2L^{-2}\right)}}$, above which an isolated WPE transitions from a stationary state to a steady walking state. Different dynamical regimes are color coded as follows: unbound clusters (yellow), stationary clusters (red), chaotic excitations (purple), particle ejection (pink), breathing mode (teal), quadrupole mode (maroon), surface oscillations (blue), compression mode (cyan), intermittent rotating mode (bright green) and rotating mode (dark green).}
\label{Fig: PS space}
\end{figure*}

\section{Active wave-particle clusters} \label{Sec: Cluster}
For a fixed value of the dimensionless wave-amplitude $R$ and the dimensionless spatial decay length $L$, an isolated WPE is stationary for small memory parameter $\tau$, and undergoes a pitchfork bifurcation at a critical memory of $$\tau_c=\sqrt{\frac{1}{R\left(1+\frac{2}{L^2}\right)}}$$ to a steady walking state~\citep{Oza2013,Valaniunsteady2021} (see Fig.~\ref{Fig: schematic}(b)). For many interacting identical WPEs with sufficiently large $L$, when an isolated WPE is either below the walking threshold or in a steady walking state above the critical memory, we find that a bound cluster of WPEs emerges. This occurs when identical WPEs are initiated in the vicinity of each other, resulting in their self-organization into a tightly bound cluster, as shown in Fig.~\ref{Fig: schematic}(c) and also observed in experiments with superwalking droplets~(see Fig.~\ref{Fig: schematic}(d) and Supplemental Video S9 of Ref.~\citep{superwalker}). The collective wave field generated by the cluster of WPEs leads to a self-generated potential well near the core of the active cluster (blue disc-shaped region) and a weak self-generated wave barrier surrounding the cluster (red annular-shaped region). The time-averaged potential, with its self-generated wave barrier and associated self-generated potential well, has some degree of similarity with certain key aspects of the bag model for hadrons~\cite{Bag-model-paper-1,Bag-model-paper-2,Bag-model-paper-3}.  In both cases, one has (i) a circular/spherical region within which the potential is approximately uniform and negative relative to the outside of the said region, with (ii) a confining potential barrier at the boundary between the interior and exterior of the bag. Some degree of parallel also exists with mean-field shell-model nuclear potentials, for the near-uniform negative potential in the cluster interior, and the repulsion-induced ``bump'' at the cluster edge \cite{cohen-nuclear-physics-book-1971}. Thus, active WPE clusters are a classical hydrodynamic system that exhibits some points of similarity with nuclear structure: the negative, quasi-uniform interior potential well corresponds to the binding of constituents, while the emergent encasing exterior barrier prevents dissociation, enabling collective excitations and long-lived, nucleus-like stability.

\begin{figure*}
\centering
\includegraphics[width=2\columnwidth]{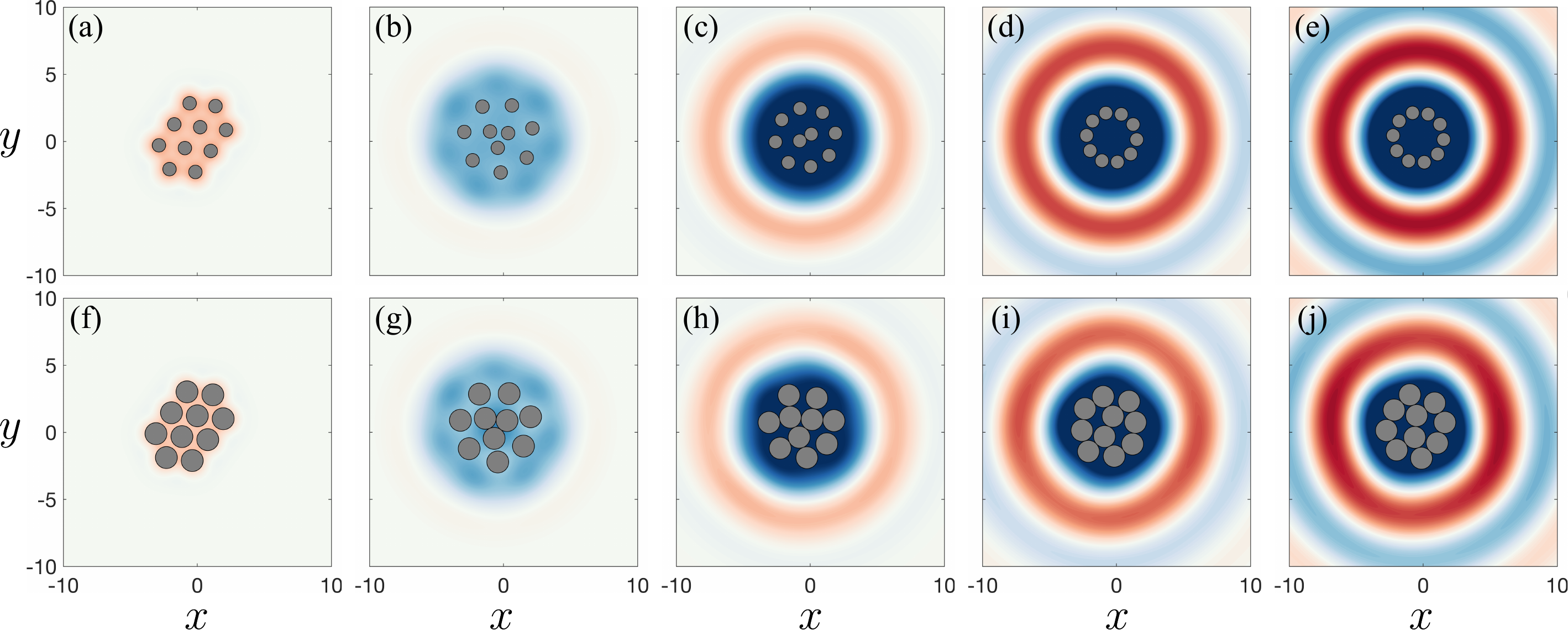}
\caption{Stationary clusters of active WPEs at low memory. A cluster of $N=10$ WPEs self-organize into various different static configurations based on the size of each particle $a$ and the length scale of spatial decay $L$ at low values of the memory parameter $\tau$. Panels (a)-(e) and (f)-(j) show configurations for two different particle sizes $a=1.01$ and $a=1.63$, respectively, for spatial decay (respective values from left to right for both top and bottom panels) $L/2\pi=0.20, 0.49, 0.78, 1.07, 1.50$. Other parameters are $\tau=0.17$ and $R=0.77$ for panels (a)-(e), and $\tau=0.44$ and $R=1.72$ for (f)-(j).}
\label{Fig: Low mem config}
\end{figure*}

Now return to Figs.~\ref{Fig: schematic}(a) and (c).  When the spatial rate of decay is sufficiently slow, for the oscillatory wave form associated with a single active particle [Fig.~\ref{Fig: schematic}(a)], it is evident from Fig.~\ref{Fig: schematic}(c) that {\em the active-particle cluster itself creates a spatially extended wave form}.  In other words, the active-particle cluster---in regimes where such clusters are stable---may itself be regarded as a wave-particle entity. Such higher-order wave-particle entities may themselves form clusters. The hierarchical ``clusters of clusters'' scenario loosely parallels the model of stable atomic nuclei as clusters formed from protons and neutrons, which are themselves considered to be three-quark clusters.  It is also worth pointing out that, if one considers an active-particle cluster with an extended wave-field as itself constituting a wave-particle entity, the ``new'' wave-particle entity can now have additional internal degrees of freedom such as the angular momentum in the center-of-momentum frame for the two or more particles that comprise the active-particle cluster.

In the next section, we show how varying the memory and wave-decay parameters induces qualitatively distinct excitation modes of the cluster, in a similar spirit to the spectrum of collective modes \cite{krane-nuclear-physics-book-1987} encountered in nuclear physics.

\section{Cluster dynamics in parameter space}\label{Sec: PS space}
In this section, we explore the dynamics of a cluster of $N=10$ WPEs, a typical cluster size formed by superwalking droplets~(see Fig.~\ref{Fig: schematic}(d)), in the parameter space formed by the spatial decay scale $L$ and the temporal-decay memory parameter $\tau$. In experiments, both the memory parameter $\tau$ and the spatial decay scale $L$ vary with forcing acceleration~\citep{Damiano2016} such that both $\tau$ and $L$ increase with forcing. Here, we explore this parameter space for a typical range of values observed in experiments. This is done for two different particle sizes: (i) $a=1.01$ (``small") -- a typical size of a walker~\citep{Couder2005WalkingDroplets,Molacek2013DropsTheory}, and (ii) $a=1.63$ (``big") -- a typical size of a superwalker~\citep{superwalker}. These two different dimensionless particle diameters of $a=1.01$ and $a=1.63$ correspond to the dimensionless wave amplitudes of $R=0.77$ and $R=1.72$, respectively. The diverse dynamical behaviors displayed by such clusters are shown in two parameter-space diagrams in Fig.~\ref{Fig: PS space}.  Each emergent class of behavior---namely stationary clusters, various forms of collective excitation, chaotic dynamics, and active-cluster decay---is detailed below.

\begin{figure*}
\centering
\includegraphics[width=1.5\columnwidth]{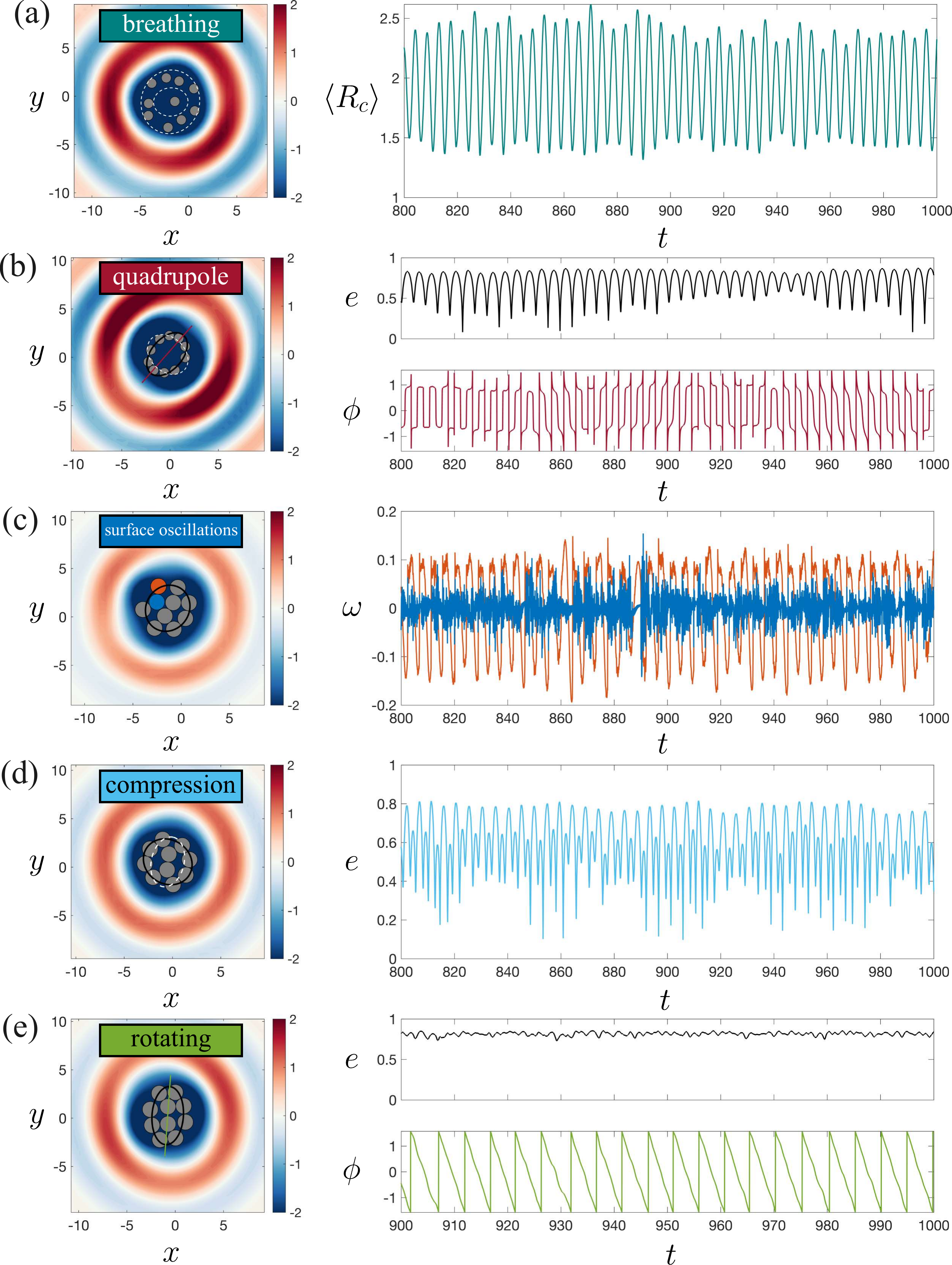}
\caption{Collective excitation modes of active WPE clusters. A cluster of $N=10$ WPEs each with size (a)-(b) $a=1.01$, $R=0.77$ and (c)-(e) $a=1.63$, $R=1.72$, display several collective excitation modes that include: (a) breathing mode (teal, $\tau=1.13$, $L/2\pi=1.5$) where the cluster stays circular (white dashed curve) and the cluster radius $\langle R_c \rangle$ varies periodically, (b) quadrupole mode (maroon, $\tau=5.91$, $L/2\pi=1.5$) where the cluster oscillates between two orthogonally oriented ellipses (black solid and white dashed ovals with maroon line indicating inclination of the ellipse), (c) surface oscillations (blue, $\tau=1.03$, $L/2\pi=0.78$) where the cluster stays circular (black solid oval) and the particles on the surface oscillate azimuthally, (d) compression mode (cyan, $\tau=1.03$, $L/2\pi=1.07$) where the cluster oscillates between a circular and an elliptical shape (black solid and white dashed ovals) and the particles on the surface undergo radial oscillations, and (e) rotating mode (dark green, $\tau=3.37$, $L/2\pi=1.07$) where the ellipse maintains a constant eccentricity (black solid oval) and its inclination (green line) rotates at a constant rate. See Supplemental Videos S1-S5~\citep{supplementary_m} for videos of modes in (a)-(e).}
\label{Fig: Cluster excitations}
\end{figure*}

\subsection{Stationary clusters}

At very small values of the memory parameter $\tau$, we obtain a static configuration of the cluster~(see Fig.~\ref{Fig: PS space}) where the arrangement of the particles in general depends on the particle size $a$, the number of particles in the cluster $N$ and (weakly) on the memory parameter $\tau$. A comparison of two different static particle arrangements for $N=10$ particles of two different sizes $a=1.01$ and $a=1.63$ is shown in Fig.~\ref{Fig: Low mem config} for different values of the spatial decay length $L$. We obtain hexagonal lattice configurations for small $L$ (rapid spatial decay) whereas for large $L$ we obtain more ring-like configurations. For the small walker-sized particle, we obtain a perfect ring configuration for large $L$. Note that both the hexagonal lattice and ring configurations have been observed in experiments with walkers at low memory~\citep{Eddielattice,Couchman_Bush_2020}. In experiments with bigger superwalking droplets, a tightly bound crystalline lattice arrangement of particles has been observed at low memory~\citep{superwalker}~(see Fig.~\ref{Fig: schematic}(d)). 

\subsection{Collective excitations}\label{Sec: collective excitations}

As the memory parameter $\tau$ is progressively increased, different excitation modes emerge in the parameter space, as shown in Fig.~\ref{Fig: PS space}, depending on particle size $a$ and spatial decay length $L$. We quantify these collective excitations by characterizing their shape dynamics. We do this by fitting an ellipse to the centers of the particles and extracting the eccentricity $e$ and inclination angle $\phi$ of the fitted ellipse as a function of time. Broadly speaking, we observe two different classes of collective excitation: (i) regular excitations that show some spatiotemporal periodicity, and (ii) irregular and chaotic excitations. Here, we explore the regular collective excitations and classify them.

\subsubsection{Breathing mode}
For the small particles of size $a=1.01$, at relatively low $\tau$, we encounter a breathing mode~(see Supplemental Video S1~\citep{supplementary_m}). In this excitation mode, the WPEs oscillate radially, as shown in Fig.~\ref{Fig: Cluster excitations}(a). The cluster remains circular but its radius,  $\langle R_c \rangle$, calculated as the mean of the radial distance of each particle from the center of the cluster, oscillates with time. This can be seen by the variation in radius of the fitted ellipse (white dashed circle) at two different times during the breathing mode. Such radial oscillations have been observed in both experiments and simulations with rings of walkers~\citep{Couchman_Bush_2020}. The breathing mode reported here parallels the monopole or “breathing” excitation of nuclei, where the nuclear surface expands and contracts uniformly in time. Such nuclear monopole vibrations, often referred to as the giant monopole resonance, correspond to coherent radial oscillations of all nucleons within the confining potential~\citep{Blaizot1980,Youngblood1999}, establishing a point of similarity with the hydrodynamic breathing oscillation observed in our WPE clusters. 

\subsubsection{Quadrupole mode}
 For the small particles of size $a=1.01$, at relatively high $\tau$ and large $L$, we find a quadrupole mode~(see Supplemental Video S2~\citep{supplementary_m}). Such a mode is seen in Fig.~\ref{Fig: Cluster excitations}(b) where the time series of the eccentricity $e$ and the inclination angle $\phi$ of the fitted ellipse are shown. In this mode, the WPEs remain in a ring structure but the ring deforms periodically between two ellipses whose inclination angle is separated by $\pi/2$ radians. The resulting dynamics bear some similarity to the nuclear quadrupole vibrational mode~\citep{krane-nuclear-physics-book-1987}. Just as quadrupole vibrations in nuclei reflect a correlated motion of nucleons within a mean-field potential, the quadrupole oscillation of the WPE cluster arises from the coupled dynamics of many active particles mediated through their shared wave field. Moreover, the periodic alternation between orthogonal axes of deformation in the active cluster mimics the nuclear case in which the quadrupole mode is associated with a characteristic $2^+$ excited state. 

\subsubsection{Surface oscillations}
For the cluster of big particles of size $a=1.63$, at relatively small $\tau$ and relatively small $L$, a collective mode appears where the particles in the cluster core remain almost stationary with small amplitude jittering whereas the particle on the cluster surface undergo angular back-and-forth oscillations with angular velocity $\omega$, as shown in Fig.~\ref{Fig: Cluster excitations}(c)~(see also Supplemental Video S3~\citep{supplementary_m}). We can see that this mode is different from the modes observed for small particles where the majority of the particles, if not all, are located at the surface. This collective mode of surface oscillations observed for bigger particles is reminiscent of surface vibrational modes in atomic nuclei, where the nucleons in the interior remain largely inactive while the surface nucleons undergo excitations, forming coherent motion along the nuclear surface~\cite{vretenar2002}.

\subsubsection{Compression mode}
For the cluster of big particles of size $a=1.63$, at relatively small $\tau$ and relatively large $L$, we obtain a compression mode where, again, particles in the core stay static while the periphery particles oscillate radially inwards and outwards~(see Supplemental Video S4~\citep{supplementary_m} and Fig.~\ref{Fig: Cluster excitations}(d)). Hence, the entire cluster oscillates between circular and elliptical shapes, as shown by black solid and white dashed ovals in Fig.~\ref{Fig: Cluster excitations}(d). This mode is analogous to the quadrupole mode observed for smaller particles, but instead of the shape oscillating between two orthogonally oriented ellipses, here, the shape oscillates between a circular and an elliptical shape. This is because the particles at the core of the cluster are almost stationary and hence the cluster can only deform along one axis. We note that, as evident from Fig.~\ref{Fig: PS space}, we do observe sensitivity to initial conditions in this regime where, depending on initial configuration, the cluster can either settle into surface oscillations (blue) or a compression mode (cyan).

\begin{figure*}
\centering
\includegraphics[width=2\columnwidth]{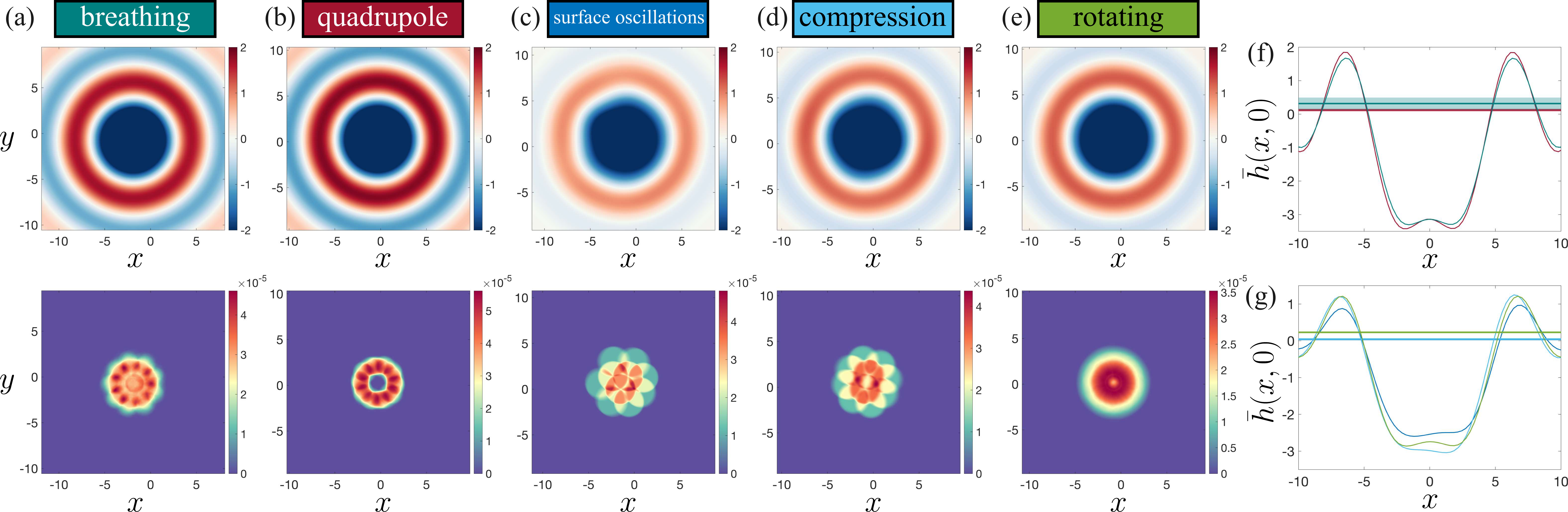}
\caption{Statistics of collective excitation modes for a cluster of $N=10$ WPEs. (a)-(e) Probability distribution showing (bottom) time-averaged particle density and (top) the time-averaged collective wave field $\bar{h}(x,y)$, corresponding to the excitation modes in Fig.~\ref{Fig: Cluster excitations}. Panel (f) shows the mean cluster kinetic energy $\bar{E}$ (solid horizontal lines) within their time-averaged collective potential $\bar{h}(x,0)$ along a horizontal centerline for different modes for small particle size in (a)-(b), whereas (g) shows the same for large particle size excitation modes in (c)-(e). Shaded regions indicate the uncertainty in the $\bar{E}$ calculated as the standard deviation. Note, also, that the color scheme for all lines in panels (f) and (g) correspond to the excitation-mode labels in panels (a)-(e).}
\label{Fig: Cluster probab}
\end{figure*}

\subsubsection{Rotating mode}
At relatively high $\tau$, we obtain a rotating mode for the bigger particles (see Supplemental Video S5~\citep{supplementary_m} and Fig.~\ref{Fig: Cluster excitations}(e)). Here, the eccentricity of the fitted ellipse stays fixed while the inclination angle decreases at a steady rate that corresponds to near-uniform angular velocity. 

This behavior is reminiscent of rotational excitation modes in atomic nuclei, specifically the rotational bands observed in deformed nuclei~\citep{krane-nuclear-physics-book-1987}. In such nuclear rotations, the nucleus possesses a permanent quadrupole deformation, and the entire nucleus rotates collectively about an axis perpendicular to its symmetry axis, maintaining its internal shape \cite{BohrMottelson1975,Ring1980,Heyde2017}. Thus, we obtain excitation modes of the collective cluster, which are reminiscent of various excitation modes in different quantum-mechanical models for nuclei, such as the shell or liquid droplet models~\citep{krane-nuclear-physics-book-1987}. 

Emergent chiral states are also a common feature in active matter, such as spontaneously rotating clusters of active droplets \cite{D1SM01795K}, electric-field–induced chiral colloidal assemblies \cite{Yan2015}, and spontaneous rotation of active nematic droplets driven by topological defects~\cite{NejadYeomans2023}.

We note that in experiments with walking and superwalking droplets, the accuracy of the simple short-range excluded-volume repulsion used in the present model remains uncertain. In particular, for larger droplets, internal deformations and shape oscillations may influence both the stability and the collective dynamics of clusters. Furthermore, in experiments with superwalkers, droplets in the core of the cluster often coalesce as the cluster grows by adding more particles. While our model treats the droplets as rigid wave-emitting particles, incorporating deformable interfaces and allowing for coalescence could introduce additional collective modes and modify stability boundaries; this is an interesting direction for future work. Nevertheless, within our framework, variations in the repulsive spring constant $K$ indicate that the qualitative nature of the collective excitation modes is largely insensitive to the precise value of $K$ (see Fig.~\ref{Fig: vary K} and Appendix~\ref{sec: diff para}).

 \begin{figure*}
\centering
\includegraphics[width=2\columnwidth]{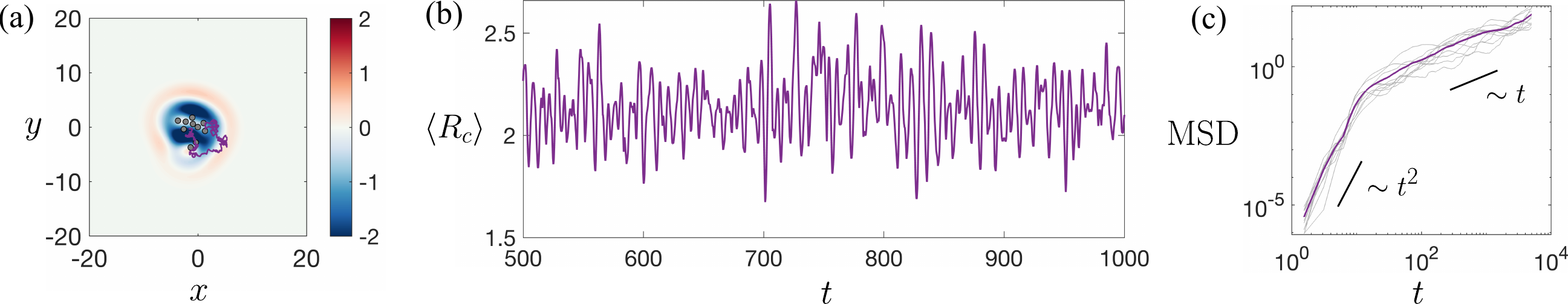}
\caption{Chaotic breathing mode of an active WPE cluster. (a) Snapshot of a cluster of $N=10$ particles in a chaotic breathing mode, alongside trajectory of the center of mass (purple). Each particle is of size $a=1.01$ with $R=0.77$, $L/2\pi=0.63$ and $\tau=2.09$~(see also Supplemental Video S6~\citep{supplementary_m}). (b) Time series of cluster radius $\langle R_c \rangle$, showing aperiodic oscillations. (c) Mean squared displacement (MSD) of the center-of-mass of the cluster, showing ballistic motion at short time scales (MSD\,$\propto t^2$) and diffusive motion at long time scales (MSD\,$\propto t$). In this panel, the individual trajectory curves are shown in gray and the averaged (over $10$ trajectories) curve is shown in purple.}
\label{Fig: Chaotic breathing}
\end{figure*}

 \begin{figure}
\centering
\includegraphics[width=\columnwidth]{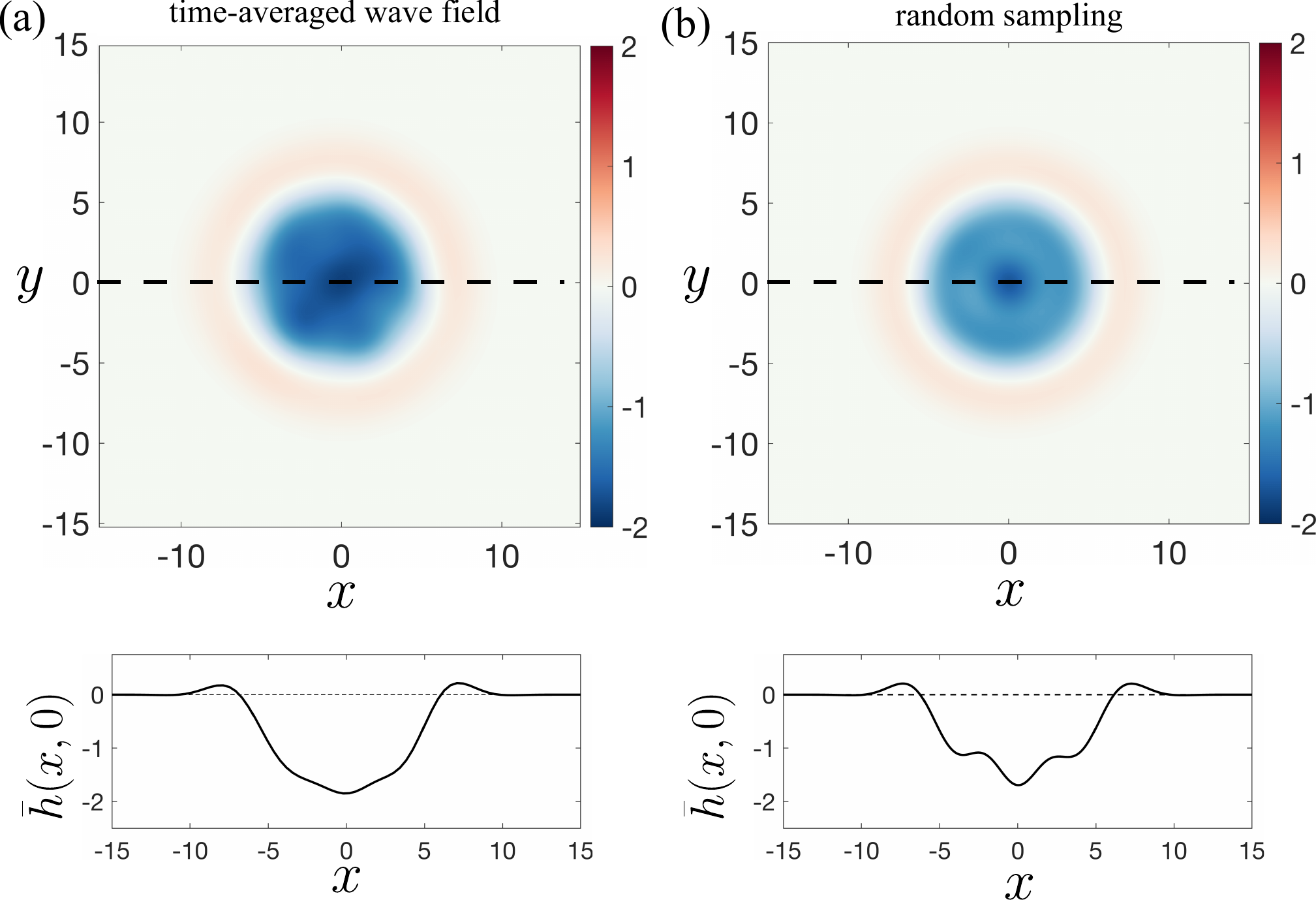}
\caption{Comparison of (a) time-averaged wave field (from $t=10$ to $t=30$) from a simulation of a chaotic breathing mode (same parameters as Fig.~\ref{Fig: Chaotic breathing}) of a cluster of $N=10$ particles with particle size $a=1.01$, with (b) the superimposed field of $N=10$ particles positioned randomly inside a disc of radius $2.14$, which is the time-averaged cluster radius in this mode that is calculated as the mean of the time series in Fig.~\ref{Fig: Chaotic breathing}(b). $100$ random samples were used to obtain the mean wave field in (b) and the particle positions were sampled uniformly inside the disc, ensuring that particles do not overlap. The wave field for each particle was taken to be that of a single bounce.}
\label{Fig: convolution}
\end{figure}

\subsection{Statistics of collective modes}

We now investigate the statistics of WPE clusters, for the collective excitation modes identified above. Figure~\ref{Fig: Cluster probab}(a)-(e) shows the probability density of particles (bottom) \footnote{The finite size of the particle as a disc of diameter $a$ is taken into account while calculating the probability density for particles.} and time-averaged collective wave field (top) for each excitation mode of both particle sizes in Fig.~\ref{Fig: Cluster excitations}. The time-averaged wave field $\bar{h}(x,y)$, which forms a time-averaged potential well, is the time average of the instantaneous collective wave field $h(x,y,t)$ formed by integrating the individual wave forms $W(\cdot)$ for all particles throughout their history. We find that although the probability density for particles is different in each mode due to the different dynamical nature of the collective excitations, the time-averaged wave field---and hence the average potential of the collective cluster---is approximately the same for all excitation modes. This behavior is reminiscent of nuclear collective models, such as the liquid-drop picture, where nucleons exhibit different vibrational or rotational motions while remaining bound by the same mean field~\citep{krane-nuclear-physics-book-1987}. The persistence of a nearly unchanged confining field across excitations can also be compared to ``bag'' models, if one views the entire nucleus as a self-consistent “bag” whose surface supports collective deformations and excitations~\citep{Bag-model-paper-1,Bag-model-paper-2,Bag-model-paper-3,DeTarDonoghue1983}. In this sense, our system has some parallels with hadronic bags at the nuclear scale, rather than bag models for quark-level confinement.

To further characterize our classical WPE cluster states, we calculate their kinetic energy
\begin{equation*}
{E(t)}=\frac{1}{N}\sum_{i=1}^{N} |\dot{\mathbf{x}}_{pi}|^2,
\end{equation*}
and time-average over the duration of the simulation to obtain the mean cluster kinetic energy $\bar{E}$. Plotting $\bar{E}$ within the time-averaged potential well (horizontal lines in Fig.~\ref{Fig: Cluster probab}(f)–(g) denote the mean value with the shaded region around them indicating the standard deviation) yields a set of discrete energy levels that are suggestive of nuclear excitation spectra~\cite{krane-nuclear-physics-book-1987}. Interestingly, some dynamical modes share nearly identical mean kinetic energies, giving rise to degeneracies that are somewhat akin to those found in nuclear shell structures, while other modes exhibit well-separated levels.

 \begin{figure*}
\centering
\includegraphics[width=2\columnwidth]{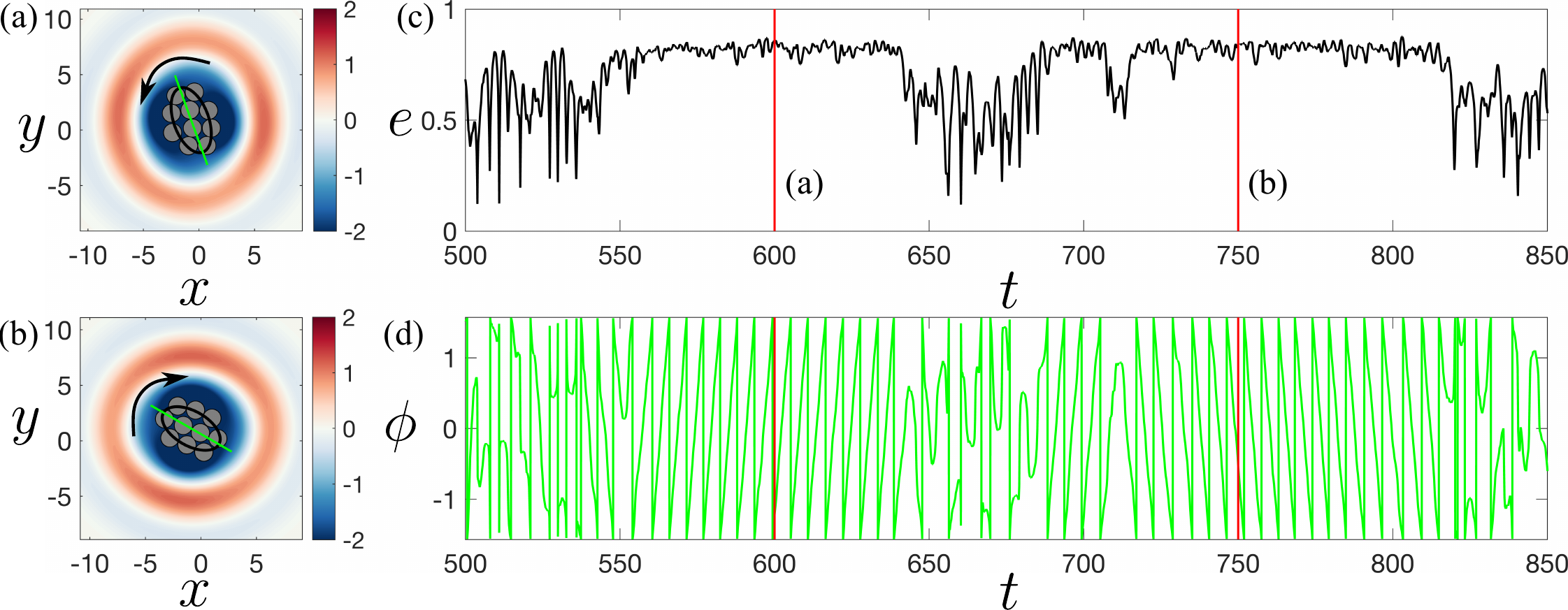}
\caption{Intermittent rotating mode of an active WPE cluster. A cluster of $N=10$ particles, each of size $a=1.63$ with $R=1.72$ and $\tau=3.08$, undergoes chaotic rotation which intermittently changes between (a) counterclockwise and (b) clockwise~(see also Supplemental Video S7~\citep{supplementary_m}). Time series of (c) eccentricity $e$ and (d) inclination angle $\phi$ of an ellipse fitted to the center of particle positions. Red lines in panels (c) and (d) correspond to the times $t=600$ and $t=750$, at which snapshots of the system are shown in (a) and (b).}
\label{Fig: Chaotic rotation}
\end{figure*}

\subsection{Chaotic dynamics}

In the parameter-space plot of Fig.~\ref{Fig: PS space}, significant regions---separating different collective excitation modes as described in Sec.~\ref{Sec: collective excitations}---are occupied by chaotic excitations. For example, in the upper parameter space, transition between the two different ``islands" of excitation modes (namely the quadrupole modes, as distinct from all other collective modes) is via a ``sea" of chaotic dynamics.  More generally, in continuous $(\tau,L)$ trajectories in either parameter space, the ``chaotic dynamics sea'' may be crossed as one moves from one patch of collective non-chaotic excitations to another. Here, we explore some interesting dynamical behaviors that are observed in the chaotic regime.

For active WPE clusters with small particles of size $a=1.01$, an example of a chaotic breathing mode is shown in Fig.~\ref{Fig: Chaotic breathing} and Supplemental Video S6~\citep{supplementary_m}.  This corresponds to a value of $\tau=2.09$ and $L/2\pi=0.63$. As shown in Fig.~\ref{Fig: Chaotic breathing}(b), we can see that in this mode the cluster radius oscillates between small and large values but it does so in an aperiodic, irregular manner. Furthermore, as shown in Fig.~\ref{Fig: Chaotic breathing}(a), we find generally that in a chaotic excitation mode, the center of mass of the cluster is not static and undergoes a random-walk-like motion (purple curve). This is quantified in Fig.~\ref{Fig: Chaotic breathing}(c) where we plot the mean-squared-displacement (MSD) averaged over an ensemble of $10$ trajectories; individual trajectory curves are shown in gray and the averaged curve is shown in purple. At shorter time scales, the center-of-mass motion is ballistic, whereas at longer time scales it becomes diffusive. Such behaviors have been reported in active droplets with a dense suspension of active particles that are confined in a droplet~\citep{PhysRevResearch.5.L032013}, where the collective rearrangements of the constituent swimmers similarly drive irregular oscillations of the droplet boundary and induce long-time diffusive wandering of the droplet’s center of mass.

In Fig.~\ref{Fig: convolution}(a), we plot the time-averaged collective wave field of the WPE cluster in the chaotic breathing mode described in Fig.~\ref{Fig: Chaotic breathing}. Similar to the collective excitation modes shown in Fig.~\ref{Fig: Cluster probab}, we again find that the time-averaged wave field retains a characteristic, reproducible profile despite the strongly fluctuating instantaneous dynamics. To test whether this structure can be captured statistically, we performed an additional calculation: we fixed the effective cluster size to the time-averaged value of the cluster radius from Fig.~\ref{Fig: Chaotic breathing}(b), randomly sampled $10$ non-overlapping particle positions within this radius, and computed the collective wave field generated by assigning each particle a single-bounce wave contribution. Repeating this procedure over $100$ independent ensembles and averaging, we obtained the wave profile shown in Fig.~\ref{Fig: convolution}(b). Despite some quantitative differences, the ensemble-averaged wave field from random sampling closely resembles the time-averaged wave field obtained from full dynamical simulations.

This comparison is in the spirit of the theorem by \citet{durey2018}, which establishes that if the chaotic dynamics of the WPE are ergodic and admit a stationary probability distribution, then the mean wave field can be expressed as the convolution of the time-integrated wave profile of a single bouncer with that stationary distribution. Our results therefore suggest that, in the chaotic regime, the cluster dynamics are indeed close to ergodic: the time-averaged wave field can be well approximated by ensemble averaging over randomly sampled particle configurations. This highlights how statistical descriptions may effectively capture the emergent collective fields of strongly fluctuating active systems.

\begin{figure*}
\centering
\includegraphics[width=2\columnwidth]{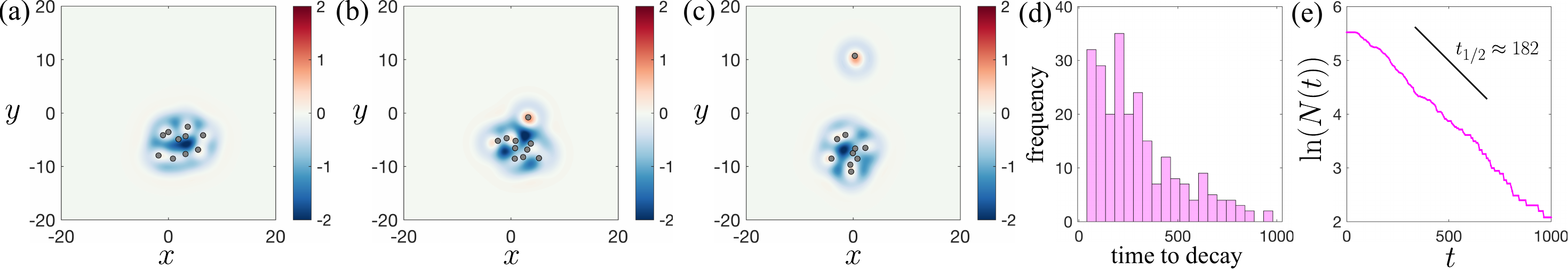}
\caption{Active WPE cluster decay statistics, for the case $N=10$. (a)-(c) Snapshots at times $t=240, 250$ and $270$, respectively, showing a first particle being ejected from the cluster. We show the distribution of (d) time of first particle decay and (e) number of undecayed clusters as a function of time. The number of undecayed clusters appears to follow an exponential decay law $N(t)=N_0 \text{e}^{-\lambda t}$ with $\lambda \approx 0.0038$ and half life $t_{1/2}=\text{ln}(2)/\lambda\approx182$. These statistics were calculated from $N_0=250$ simulations with different random initial perturbations in particle position. Parameter values were fixed to memory parameter $\tau=1.77$, decay length scale $L/2\pi=0.49$,  particle size $a=1.01$ and $R=0.77$.}
\label{Fig: cluster decay}
\end{figure*}

For active WPE clusters with big particles of size $a=1.63$, one observes in the parameter space shown in the lower part of Fig.~\ref{Fig: PS space} that near the boundary separating chaotic excitations (purple) and the rotating mode (dark green), there is an intermittently chaotic rotating mode (bright green). As shown in Fig.~\ref{Fig: Chaotic rotation} and Supplemental Video S7~\citep{supplementary_m}, in this intermittent rotating mode, the cluster dynamics intermittently switches between clockwise and counterclockwise rotations. This can be seen in the time series of the eccentricity and the inclination angle of the ellipse fitted to the cluster, as shown in Fig.~\ref{Fig: Chaotic rotation}(c) and (d), respectively. During the rotating regions of the time series, the eccentricity is almost constant and the inclination angle varies almost linearly. The time series is interspersed with rotating regions of both chiralities, i.e.~both clockwise and counterclockwise. Such intermittency is reminiscent of dynamical regimes in which the system explores a superposition of competing modes rather than locking into a single stable one. In published work on hydrodynamic quantum analogs, similar intermittent dynamics between different periodic motions have been observed for a walking droplet in a harmonic potential~\citep{PhysRevLett.113.104101}. Here, the discrete spectrum of periodic motion of the walker is related to eigenstates and the intermittent switching between them is related to the superposition of eigenstates. In a similar spirit, the intermittent switching of our cluster between clockwise and counterclockwise rotations can be interpreted as a superposition of the two chiral rotational states, with the system transiently exploring both nearly degenerate modes.

\subsection{Active-cluster decay}

In the parameter space of Fig.~\ref{Fig: PS space}, for very small values of the spatial decay length $L$, clusters do not form and the particles do not stay bound. Near the boundary between this unbound-state regime (yellow) and the chaotic excitation regime (purple), we have a small region of the parameter space where we observe single-particle ejections (pink). Typically, after the single-particle ejection the remaining cluster eventually becomes unbound on a longer time scale. However, in this regime, a single particle always gets ejected first, as illustrated in Fig.~\ref{Fig: cluster decay}(a)-(c). We note that such single-particle ejections from a cluster are ubiquitously observed in experiments with clusters of superwalking droplets~\citep{superwalker}, where the cluster is chaotically vibrating at large memory parameter and is on the verge of disintegration.

To quantify the statistics of cluster decay, we plot the distribution of ejection times for the first particle from the cluster (Fig.~\ref{Fig: cluster decay}(d)), as well as the number of non-decayed clusters as a function of time (Fig.~\ref{Fig: cluster decay}(e)). As shown in Fig.~\ref{Fig: cluster decay}(d), the distribution of first-particle ejection times exhibits a decaying nature, indicating that it is progressively less likely for a particle to remain in the cluster for very long before being ejected. Figure~\ref{Fig: cluster decay}(e) shows the time series of the logarithm of the number of non-decayed clusters, $N(t)$, starting from an initial $N_0=250$ clusters, i.e., 250 different initial conditions of the system. Interestingly, $N(t)$ follows an exponential decay law of the form 
\begin{equation*}
N(t)=N_0\, \text{e}^{-\lambda t},    
\end{equation*}
with $\lambda\approx 0.0038$. This decay law is identical in mathematical form to that for radioactive nuclear decay, where the probability of decay also has an exponential form, and the system exhibits a well defined half-life. In our case, the half-life of the active-cluster decay can be calculated as $t_{1/2}=\ln(2)/\lambda\approx182$. This point of similarity suggests that the chaotic ejection of particles from the cluster can be interpreted as a memoryless Poisson process, just as nuclear decay is governed by the probabilistic nature of unstable nuclei, thereby highlighting similarities between the emergent statistics of active-matter clusters and fundamental decay processes in quantum systems. In addition to complete particle ejection, we also find weakly bound particles orbiting the cluster after ejection (see Fig.~\ref{Fig: Halo Nuclei} and Appendix~\ref{sec: diff para}); this has some similarities to halo nuclei~\citep{Ye2025}. 


\section{Conclusion}\label{Sec: DC}

We showed that multiple classical wave–particle entities (WPEs) can self-organize into stable, nucleus-like structures that transition from a static state to a range of collective excitation modes as the system memory increases. These include breathing, quadrupole, compression, and rotational excitations, which arise from the interplay between inertia, dissipation, and long-lived nonlinear wave fields of the WPEs. Despite their dynamical diversity, the distinct excitation modes give rise to a common time-averaged collective potential, enabling direct parallels with mean-field and bag models in nuclear physics. In certain parameter regimes, WPE clusters can spontaneously emit individual particles, with decay statistics that appear to follow an exponential law; this is reminiscent of radioactive decay processes. These findings demonstrate that memory-mediated active systems can display emergent confinement, discrete excitation spectra, and stochastic decay, all features typically associated with bound quantum systems. In this sense, active WPE clusters serve as a classical model for exploring the interplay between deterministic dynamics and probabilistic emergent behavior in non-equilibrium systems.

Our work opens avenues to explore in detail the possibility of hydrodynamic nuclear analogs in both experiments and simulations with WPEs. In particular, it would be interesting to investigate (i) analogs of stable and unstable nuclei by examining the stability of differently sized WPE clusters, (ii) active WPE clusters with two species to model proton–neutron or positive–negative–charge analogs, and (iii) possible analogs of nuclear scattering. These and other points of interest for future explorations, motivated by our findings, are discussed in Appendix \ref{sec:ExtendedDiscussion}.

\section*{Acknowledgments}
R.V.~acknowledges the support of the Leverhulme Trust [Grant No. LIP-2020-014] and the ERC Advanced Grant ActBio (funded as UKRI Frontier Research Grant EP/Y033981/1).

\appendix

\begin{figure*}
\centering
\includegraphics[width=1.5\columnwidth]{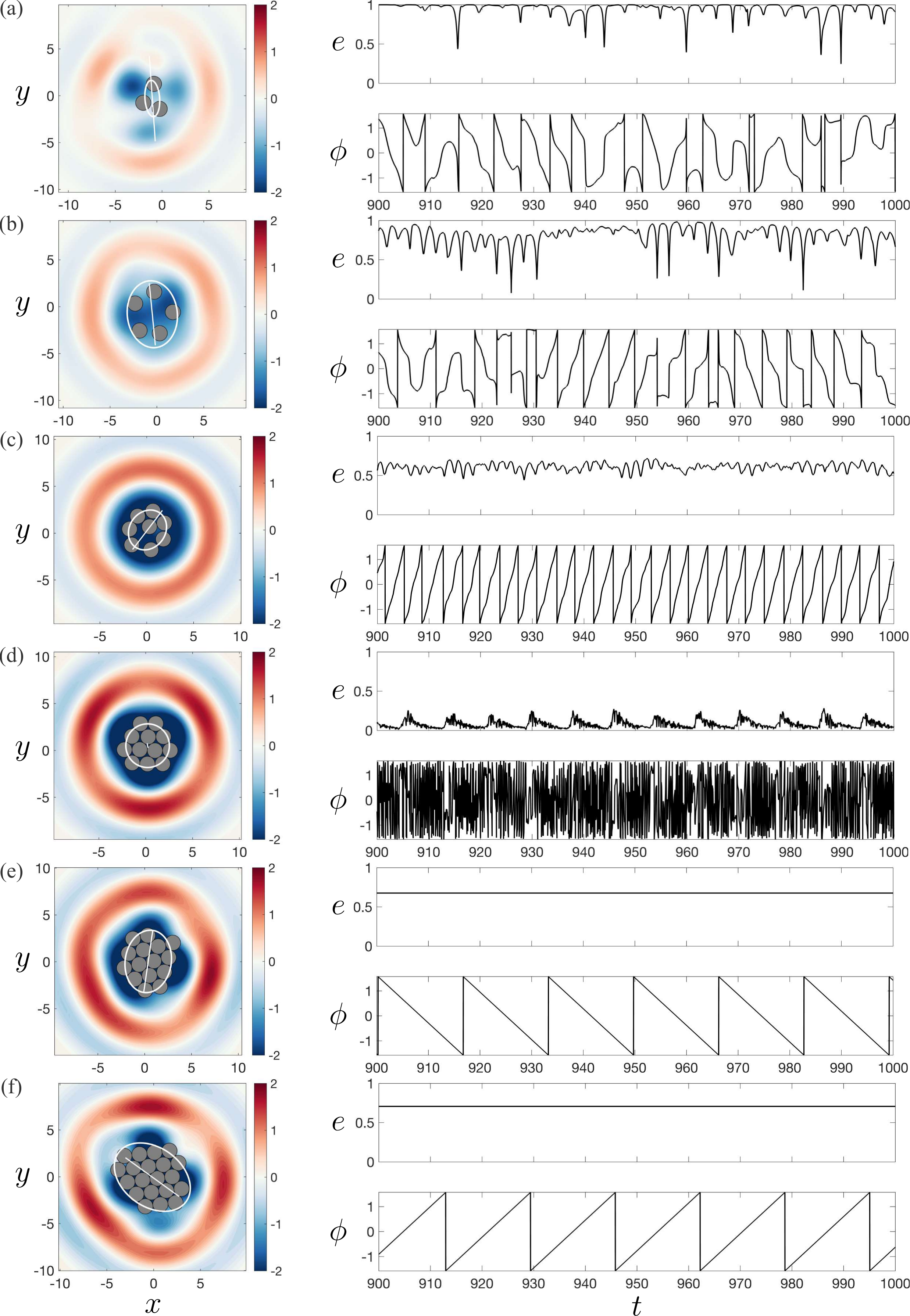}
\caption{Cluster dynamics with variations in the number of particles $N$. For fixed parameter values $\tau=3.37$, $L/2\pi=1.07, a=1.63, R=1.72$, a cluster of $N=10$ WPEs undergoes a rotation-mode excitation as shown in Fig.~\ref{Fig: Cluster excitations}(e). For these same parameter values, we plot a snapshot of the particles along with their wavefields, time series of eccentricity $e$ and the inclination angle $\phi$ of the fitted ellipse for (a) $N=3$, (b) $N=5$, (c) $N=8$, (d) $N=12$, (e) $N=15$ and (f) $N=20$ particles. White oval curves show the fitted ellipse with the enclosed white straight line corresponding to the direction of the major axis of the ellipse. See Supplemental Videos S9-S14~\citep{supplementary_m} for videos of modes in (a)-(f).}
\label{Fig: vary N}
\end{figure*}

\begin{figure*}
\centering
\includegraphics[width=1.5\columnwidth]{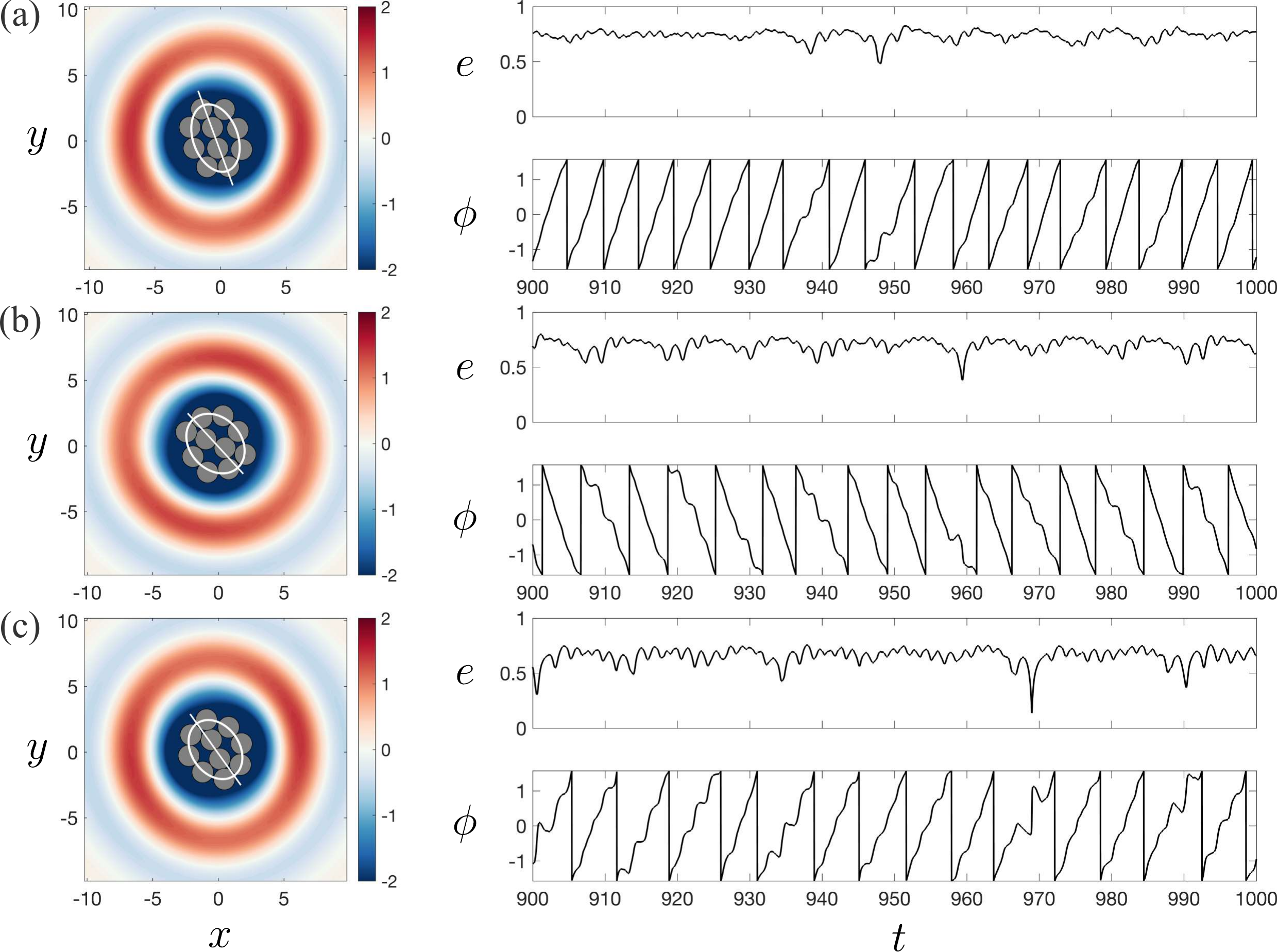}
\caption{Cluster dynamics with variations in the interaction spring constant $K$. For fixed parameter values $\tau=3.37$, $L/2\pi=1.07, a=1.63, R=1.72, K=1000$, a cluster of $N=10$ WPEs undergoes a rotation-mode excitation as shown in Fig.~\ref{Fig: Cluster excitations}(e). For these same parameter values, we plot a snapshot of the particles along with their wavefields, time series of eccentricity $e$ and the inclination angle $\phi$ of the fitted ellipse for (a) $K=500$, (b) $K=100$ and (c) $K=50$. White oval curves show the fitted ellipse with the enclosed white straight line corresponding to the direction of the major axis of the ellipse.}
\label{Fig: vary K}
\end{figure*}

\section{Numerical implementation details}\label{App 1}

We solve the integrodifferential trajectory equations for $N$ WPEs using a modified Euler method. The system is initiated at $t=0$ with the particle position of the $i$th particle is
\begin{equation}
\mathbf{x}=(x_{pi},y_{pi}),
\end{equation}
and all the particle positions are initialized randomly in a circle of radius $a\sqrt{N/2}$, while making sure that they are not overlapping. The new position of the $i$th particle for $t>0$ is calculated from its old position using a forward Euler step 
\begin{align*}
x_{pi}(t_{n+1}) &= x_{pi}(t_n)+\Delta t 
\,u_{pi}(t_n)
\\
y_{pi}(t_{n+1}) &= y_{pi}(t_n)+\Delta t\,v_{pi}(t_n),
\end{align*}
where 
\begin{equation}
\mathbf{\dot{x}}=(u_{pi},v_{pi}) 
\end{equation}
is the $i$th particle velocity. To calculate the new velocity, we use the new position and a backward Euler step via
\begin{widetext}
\begin{align}\label{eq:nm1}
u_{pi}(t_{n+1})&=u_{pi}(t_n)+{\Delta t}\Biggl[-u_{pi}(t_{n+1}) - R\int_{0}^{t_n} {W'(|\mathbf{x}_{pi}(t_{n+1})-\mathbf{x}_{pi}(s)|)}\frac{x_{pi}(t_{n+1})-x_{pi}(s)}{|\mathbf{x}_{pi}(t_{n+1})-\mathbf{x}_{pi}(s)|}\,e^{-(t_{n+1}-s)/\tau}\,ds\notag\\
&- R \sum_{\substack{j=1 \\ j\neq i}}^{N} \int_{0}^{t_n} {W'(|\mathbf{x}_{pi}(t_{n+1})-\mathbf{x}_{pj}(s)|)}\frac{x_{pi}(t_{n+1})-x_{pj}(s)}{|\mathbf{x}_{pi}(t_{n+1})-\mathbf{x}_{pj}(s)|}\,e^{-(t_{n+1}-s)/\tau}\,ds\ + \sum_{\substack{j=1 \\ j\neq i}}^{N} F^{rep}_{xij} \Biggr]
\end{align}
\begin{align}\label{eq:nm2}
v_{pi}(t_{n+1})&=v_{pi}(t_n)+{\Delta t}\Biggl[-v_{pi}(t_{n+1}) - R\int_{0}^{t_n} {W'(|\mathbf{x}_{pi}(t_{n+1})-\mathbf{x}_{pi}(s)|)}\frac{y_{pi}(t_{n+1})-y_{pi}(s)}{|\mathbf{x}_{pi}(t_{n+1})-\mathbf{x}_{pi}(s)|}\,e^{-(t_{n+1}-s)/\tau}\,ds\notag\\
&- R \sum_{\substack{j=1 \\ j\neq i}}^{N} \int_{0}^{t_n} {W'(|\mathbf{x}_{pi}(t_{n+1})-\mathbf{x}_{pj}(s)|)}\frac{y_{pi}(t_{n+1})-y_{pj}(s)}{|\mathbf{x}_{pi}(t_{n+1})-\mathbf{x}_{pj}(s)|}\,e^{-(t_{n+1}-s)/\tau}\,ds\ + \sum_{\substack{j=1 \\ j\neq i}}^{N} F^{rep}_{yij} \Biggr],
\end{align}
where
\begin{equation*}
F^{rep}_{xij} = 
\Big{\{}
    \begin{array}{lr}
        K\frac{{x}_{pi}(t_{n+1})-{x}_{pj}(t_{n+1})}{|\mathbf{x}_{pi}(t_{n+1})-\mathbf{x}_{pj}(t_{n+1})|}\left(a - |\mathbf{x}_{pi}(t_{n+1})-\mathbf{x}_{pj}(t_{n+1})|\right), \,\,\,\,\,\,\,\,\,\,\,\,\,\,\,\,& \text{if } |\mathbf{x}_{pi}(t_{n+1})-\mathbf{x}_{pj}(t_{n+1})|<a\\ 
        0, & \text{otherwise}
    \end{array}
\end{equation*}
\begin{equation*}
F^{rep}_{yij} = 
\Big{\{}
    \begin{array}{lr}
        K\frac{{y}_{pi}(t_{n+1})-{y}_{pj}(t_{n+1})}{|\mathbf{x}_{pi}(t_{n+1})-\mathbf{x}_{pj}(t_{n+1})|}\left(a - |\mathbf{x}_{pi}(t_{n+1})-\mathbf{x}_{pj}(t_{n+1})|\right), \,\,\,\,\,\,\,\,\,\,\,\,\,\,\,\,& \text{if } |\mathbf{x}_{pi}(t_{n+1})-\mathbf{x}_{pj}(t_{n+1})|<a\\
        0, & \text{otherwise.}
    \end{array}
\end{equation*}
\end{widetext}
 We use a numerical time step of $\Delta t=2^{-6}$ for integration. The integrals in Eqs.~\eqref{eq:nm1} and \eqref{eq:nm2} were performed using the trapezoidal rule, where we consider the contribution from all the previous impacts for the first 1280 timesteps relative to memory ($t/\tau=20$) and then the contribution from the last 1280 impacts for $t/\tau>20$. At 1280 previous impacts, the exponential damping factor has reached $e^{-20} \approx 10^{-9}$ so we neglect all contributions from impacts beyond 1280 previous steps.

\section{Additional results on cluster dynamics for varying parameters}\label{sec: diff para}

In the main text, we considered the dynamics of active WPE clusters with respect to variations in the dimensionless wave memory parameter $\tau$, the dimensionless spatial decay length scale $L$, and the dimensionless particle size $a$ (which fixes the value of the dimensionless wave amplitude $R$). In this appendix, we explore the effects of varying other system parameters, namely (i) the number of particles $N$ and (ii) the interaction spring constant $K$; we then explore (iii) an interesting ``halo cluster'' effect, which emerges for a larger particle size than that employed in the main text of the paper.

Figure~\ref{Fig: vary N} shows how the cluster dynamics can change with respect to variations in the number of particles $N$ in the cluster. We have fixed the other parameters corresponding to the rotation mode shown in Fig.~\ref{Fig: Cluster excitations}(e) for $N=10$. Firstly, we note that changing the number of particles from $N=3$ to $N=20$, we still obtain bound clusters with dynamic excitations. This suggests that active WPE clusters are stable across a wide range of particle sizes $N$, as observed in experiments with superwalking droplets~\citep{superwalker}. We find however, that the particular excitation modes vary significantly with changes in $N$. For $N=10$ we have the rotation mode, and upon slightly reducing the particle number to $N=8$, we still find a qualitatively similar rotation mode as shown in Fig.~\ref{Fig: vary N}(c) and Supplemental Video S11. Further decreasing the particle number to $N=5$ and $N=3$, we obtain chaotic excitations, as shown in Fig.~\ref{Fig: vary N}(a)-(b) and Supplemental Videos S9-S10. Increasing to $N=12$, we find a qualitatively different arrangement of particles in the cluster, which now has three-fold symmetry. Here, we still obtain periodic rearrangements of the cluster (see Fig.~\ref{Fig: vary N}(d) and Supplemental Video S12) but the mode cannot be well described by the fitted ellipse since it does not have dipolar symmetry. Further increasing particle numbers to $N=15$ and $N=20$, we again obtain a rotation mode, but this rotation mode is slightly different from the one at $N=10$ since here we have rigid-body rotation of the whole cluster (see Fig.~\ref{Fig: vary N}(e)-(f) and Supplemental Videos S13-14).  We highlight the emergence of crystalline ordering when the particle number $N$ is sufficiently large: for the lower particle numbers in Figs.~\ref{Fig: vary N}(a)-(c) the particles in the cluster are loosely packed, whereas for the larger particle numbers in Figs.~\ref{Fig: vary N}(d)-(f) they exhibit hexagonal close packing \cite{KittelBook} (cf.~Fig.~\ref{Fig: schematic}(d)).  This transition to crystalline ordering at sufficiently large $N$ may be viewed as the self-generated mean-field potential becoming sufficiently ``deep and narrow'' for the active particles to become close packed (for the parameters considered in the present simulations). Interestingly, the ordered rotating clusters we observe have similarities with recent observations of chiral crystals formed by self-spinning living units. For example, starfish embryos can spontaneously assemble into chiral crystals that exhibit collective rotational order~\cite{Tan2022OddDynamics}. Such systems illustrate how active units can self-organize into chiral lattices with robust rotational dynamics, hinting that our WPE clusters may likewise serve as minimal models for exploring active chiral crystallinity in synthetic systems.


Figure~\ref{Fig: vary K} shows how the cluster dynamics vary with the dimensionless spring constant $K$, which controls the short-range repulsive interactions between particles. We fix the parameters corresponding to a rotation-mode excitation at $K = 1000$ (see Fig.~\ref{Fig: Cluster excitations}(e)) and progressively decrease $K$, thereby softening the repulsive interactions. As shown in Fig.~\ref{Fig: vary K}, the rotating mode persists as $K$ is reduced; however, for smaller $K$ values, the time evolution of the inclination angle of the fitted ellipse develops oscillations, indicating a wobbling motion of the cluster arising from the weakened repulsive coupling between particles. This suggests that the cluster dynamics are largely insensitive to variations in $K$; the rotational mode remains robust even when $K$ is varied by two orders of magnitude.

\begin{figure*}
\centering
\includegraphics[width=2\columnwidth]{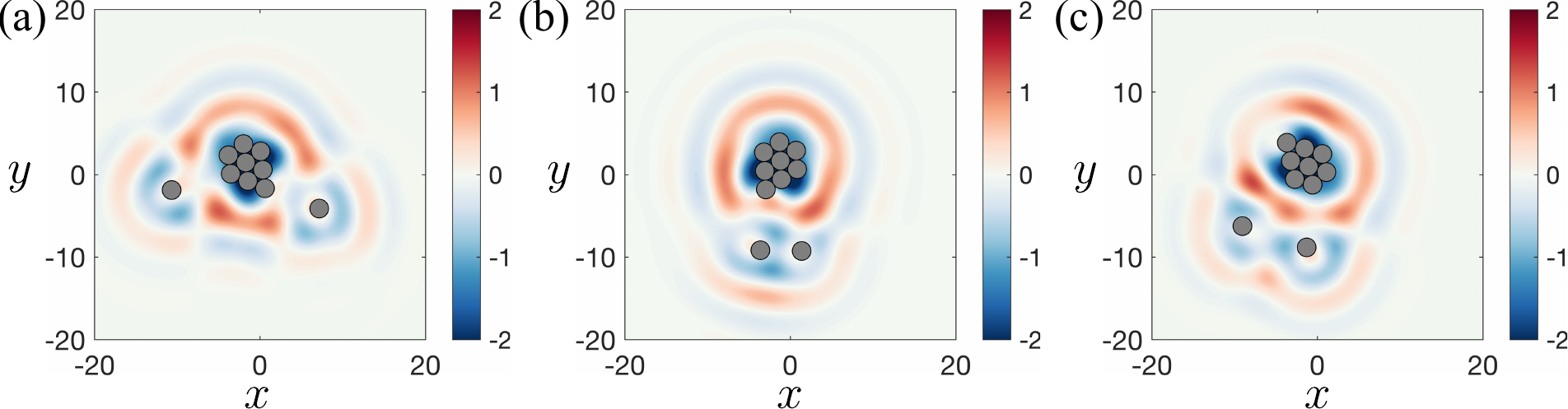}
\caption{Decay of an active wave-particle cluster with the ejected particles bound to a ``halo'' after ejection. A cluster of $N=10$ WPEs of particle size $a=2.26$ undergoes chaotic oscillations and ejects particles which remain weakly bound to the cluster~(see also Supplemental Video S15). Snapshots are shown at (a) $t=240$, (b) $t=270$, and (c) $t=300$. Other parameters are fixed to $R=2.86$, $L/2\pi=1.07$ and $\tau=8.64$.}
\label{Fig: Halo Nuclei}
\end{figure*}

In addition to active wave-particle clusters completely ejecting a particle [see Fig.~\ref{Fig: cluster decay}(a)-(c)], for larger particle size, we find regimes where the ejected particles stay bound to the cluster but cannot penetrate back inside the cluster due to the wave barrier.  This latter effect, as shown in Fig.~\ref{Fig: Halo Nuclei} and Supplemental Video S15, exhibits some degree of similarity with a halo nucleus wherein a core nucleus is surrounded by one or more orbiting protons or neutrons~\citep{Ye2025}. This observation leads to the following closely related idea.  At several points in the main text, we noted certain similarities between the bag model of hadrons, and the confining potential barrier that develops at the edges of stable active-particle clusters.  We also saw, that in the case where the oscillatory wave-form associated with a single active particle decays sufficiently slowly in space, the resulting ``bag'' that confines a stable active-particle cluster may develop a series of layers -- see (i) the annular emergent-waveform potential well that is shaded blue, which lies {\em outside} the self-generated potential barrier shaded red, in Fig.~\ref{Fig: schematic}(c), together with (ii) the more strongly layered emergent waveforms in Figs.~\ref{Fig: Low mem config}(e,j), Figs.~\ref{Fig: Cluster probab}(a)-(b), and Fig.~\ref{Fig: Halo Nuclei}. These emergent {\em layered bags} consist of alternating wall-like potential barriers that are separated by potential wells. Such layered bags are directly related to the ``halo cluster'' effect described earlier in this paragraph, since an active particle ejected from the core of an active-particle cluster may become trapped (for a period) in the ``moat'' between an adjacent pair of layered potential-wall maxima. 

\section{Extended discussion}\label{sec:ExtendedDiscussion}
To augment the brief discussion in Sec.~\ref{Sec: DC}, we here list several possible avenues for future work:
\begin{enumerate}
    \item  It would be useful to investigate whether there exists an active-particle-cluster version of the {\em semi-empirical mass formula} of nuclear physics  \cite{krane-nuclear-physics-book-1987}. Such a formula might have the form 
    \begin{equation}
    \mathcal{B}\approx A N - B \sqrt{N}+\cdots, 
    \end{equation}
    where $\mathcal{B}$ is the binding energy of an active-particle cluster, $AN$ is a bulk term proportional to the number of particles $N$ in the cluster (with $A > 0$), $-B\sqrt{N}$ is a surface term proportional to the number of particles along the perimeter of the cluster (with $B > 0$), and additional terms may be employed to model effects such as particle-particle repulsion.  Perhaps such a formula could be used to determine a condition for an active-particle cluster to be stable, e.g.~to determine the $N$-dependent boundary between (i) the ``unbound cluster'' region that is shaded yellow in the parameter-space maps in Fig.~\ref{Fig: PS space}, and (ii) all other regions of the parameter space (which correspond to bound clusters of various forms).
    \item We only considered one type of active particle, in modeling active wave-particle clusters.  However, the governing system of equations [Eqs.~(\ref{eq: dimensional_2D}) and (\ref{eq: dimensional_multi}) in the main text] could be easily generalized to two or more active-particle types.  In particular, if {\em two types of active particle were to be employed}---with one (proton like) type having long-range Coulomb repulsion to augment the short-range repulsion, and the other (neutron like) type only having short-range repulsion---then certain potential parallels with nuclear physics (or a lack thereof) might become clearer. Adding a third spatial dimension to the model may also be clarifying, in this regard.  In adding the third spatial dimension, we would of course no longer have a realistic model for bouncing droplets, but rather would seek in this context a possible active-particle nuclear analog that is inspired by---but not constrained by---the physics of bouncing droplets, in a similar spirit to the recent extension of the walking-droplet model to three dimensions~\citep{doi:10.1098/rspa.2024.0986}.
    \item It would be interesting to study the {\em scattering of active-particle clusters from one another}, together with the {\em scattering of one fast active particle from an active-particle cluster}.  Such studies could lead to an analog notion of differential and total scattering cross sections, together with associated concepts such as scattering resonances.
    \item Do {\em active-particle-cluster analogs of nuclear fission and nuclear fusion} exist?  If an active-particle-cluster analog of nuclear fission exists, do strongly-deforming collective excitations play a role in the fission mechanism (as least for some of the fission decay channels)?
    \item It would be interesting to study a gas of active-particle clusters, since the average wave barrier surrounding each cluster implies they will repel one another when they touch (unless they collide with sufficiently high speed to initiate a fusion event).
    \item Does the center-of-mass of an active-particle cluster {\em recoil} when a particle is ejected [cf.~Fig.~\ref{Fig: cluster decay}(a)-(c)]?  If such a recoil velocity exists, over what timescale is its associated drift erased by the diffusive motion evident in Fig.~\ref{Fig: Chaotic breathing}?  Presumably, the time evolution of the probability density for the center-of-mass location, following ejection of a particle from an active-particle cluster, may be modeled using a Fokker-Planck or Kramers-Moyal equation \cite{Risken1989}.
    \item Can a {\em crystal lattice of active-particle clusters} be formed?  If such a crystal lattice can indeed be formed, does it exhibit any degree of similarity with the M\"{o}ssbauer effect \cite{krane-nuclear-physics-book-1987} when one of the active-particle clusters ejects a particle?  In particular, when one of the clusters ejects a particle, does the entire lattice recoil or just the emitting active-particle cluster?
    \item  The mean particle kinetic energy, relative to the center-of-mass of the active-particle cluster, may be considered as directly proportional to an associated pseudotemperature (namely an ``effective temperature'' or ``characteristic temperature'' that is not a true thermodynamic temperature since the active-particle cluster is not in thermal equilibrium \cite{PellicciaPaganin2025}). Moreover, we may adapt language in the bag-model paper of \citet{Bag-model-paper-2}, by likening our active-particle clusters to an active-particle gas bubble that is contained within a ``liquid'' of approximately constant pressure. Under this view, we can think of the active-particle cluster as being confined via the external pressure that is mentioned in the previous sentence; the cluster radius is then determined by balancing the cluster-expanding force associated with the cluster pseudotemperature, against the cluster-contracting force of the external-``liquid'' pressure.  In this regard, a key parameter is the active-particle-cluster version of the ``bag constant'', which for our model is given by the mean-field potential per unit area, in the cluster interior (this may be estimated via the same random-sampling method used to obtain Fig.~\ref{Fig: convolution}(b)). 
    \item How does the pseudotemperature behave in the temporal vicinity of a particle-ejection event?  For example, it would be interesting to study whether or not the pseudotemperature increases immediately before an ejection event, and if it has the property analogous to evaporative cooling of becoming reduced after an ejection event.
    \item The oscillation frequency associated with our active particles bears some degree of similarity with the Zitterbewegung frequency \cite{sakurai_1961,DarrowBush2025} that naturally occurs in relativistic field theories.  Since antiparticles arise inevitably in such theories, {\em can antiparticle analogs be found in our model?}  In particular, does reversing the sign of the waveform $W$ create an active-particle analog of an antiparticle? If this is indeed the case, then it would be interesting to study possible analogs of a bound proton-antiproton pair (protonium) or a bound electron-positron pair (positronium) using active-particle pairs that have equal masses but a waveform of opposite sign.  Note that in both of these cases we would need to augment the interparticle force $F_{ij}^{int}$ with a Coulomb-type term.  Can such a model give an analog for particle-antiparticle pair creation and pair annihilation?  Alternatively, perhaps a Coulomb term could be used to model particles of opposite charge, with or without the use of a change in sign for $W$.  Could an analog of a hydrogenic atom be modeled along these lines, using active particles of very different masses together with a Coulomb term corresponding to particles with opposite electric charge? If such a hydrogenic-atom active-particle analog de-excites from a higher-energy state to a lower-energy state, is the resulting energy radiatively carried away in a suitable photon-analog structure that is contained in the wave field?
    \item Our governing nonlinear differential equation, namely Eqs.~(\ref{eq: dimensional_2D}) and (\ref{eq: dimensional_multi}) in the main text, is ultimately derived from a suitable coarse graining of a finite-order underpinning nonlinear differential equation (namely the Navier-Stokes equations of classical fluid mechanics)~\citep{molacek_bush_2013,Molacek2013DropsTheory,Oza2013}.  Thus the WPEs in our model may be viewed as oscillating soliton-like \cite{kivshar-agrawal-book} solutions (breathers \cite{akhmediev2001}) of a governing nonlinear differential equation.  This immediately leads to the observation that the key ideas of the present paper might be adapted to study WPEs associated with soliton breather solutions arising from other nonlinear finite-order partial differential equations.  Examples include the WPEs known as ``oscillons'' that arise in periodically driven granular media \cite{Oscillons1996a,Oscillons1996b,Oscillons1997,Oscillons2006}, together with the closely related concept of soliton-breather gases \cite{ElTovbis2021}.  Regarding the former example, it is of relevance in the present context that an early experimental study showed that when solitonic breather-like states have opposite phase---namely fast-oscillation phases that differ by $\pi$ radians---the associated structures behave analogously to particle--antiparticle pairs in the sense that ``Oscillons of like phase show a short-range repulsive interaction, while oscillons of opposite phase attract and bind'' \cite{Oscillons1996a}. These remarks are suggestive of a general approach, whereby an underlying finite-order nonlinear partial differential equation which admits breather-like solitons (WPEs) can then lead to an analog of Eqs.~(\ref{eq: dimensional_2D}) and (\ref{eq: dimensional_multi}) {\em for those solitons} (WPEs), namely an emergent integrodifferential equation governing WPE motion. This program is conceptually close to the causal ``theory of the Double Solution'' due to  de Broglie, with what we term the wave form $W$ being associated with ``a continuous $\Psi$ solution'' and what we term the active particle being associated with ``another solution ... having an amplitude singularity that is in general mobile'' which ``would obey a non-linear wave equation'' (see pp.~216-218 of the book by \citet{de-broglie-book-1960}).  Indeed, such a program may be viewed as motivating from whence the ``double solution'' actually emerges~\footnote{Page 289 of the book by \citet{de-broglie-book-1960} suggests---in the context of the ``double solution'' model---that ``a piling up of [a] great number of singular regions'' which are ``very close together or even overlapping might permit the correct forecasting of certain nuclear phenomena''.  This conjecture bears some conceptual similarity to the core study of the present paper.}.

\end{enumerate}

We close with some speculations. It is natural to consider a broader context provided by parallels that have been drawn, between classical active wave-particle entities and certain model quantum-mechanical systems~\citep{Bush2015,Bush_2021,Bush2024}.  In particular, researchers have investigated the possibility that the former, classical, systems may exhibit behaviors hitherto thought exclusive to the quantum realm. In a similar vein, 't Hooft poses the logical possibility that quantum mechanics is a mean-field theory that may emerge---without needing any modification to the existing formalism---via coarse-graining of an as-yet unknown {\em classical} system such as a suitable cellular-automaton \cite{hooft-cellular-automaton-book}. A key idea, in this regard, is that such underlying classical systems are worth further investigation since they may not be necessarily ruled out by standard ``no go'' arguments associated with Bell inequalities and hidden-variable theories \cite{hooft-cellular-automaton-book}.  While we are unable at this stage to substantiate any claim regarding the applicability of our findings to the genuinely quantum-mechanical domain of nuclear and composite-particle physics, it is nevertheless interesting to further explore the extent to which other nuclear-physics and composite-particle phenomena may have direct or partial analogs with entirely-classical clusters of wave-particle entities.  In pursuing such a line of inquiry, it appears likely that a different form for the wave form $W$ may need to be employed.  {\em Can a suitably modified form of the classical integrodifferential expressions in Eqs.~(\ref{eq: dimensional_2D}) and (\ref{eq: dimensional_multi}) of the main text, with a suitably modified wave form and a suitably modified repulsive term, be amenable to coarse graining which subsequently leads to a formalism that more closely approximates an emergent quantum system?  Can the integrodifferential equation itself be obtained via breathing-mode solitonic solutions to an underpinning finite-order partial differential equation?}

\bibliography{wave_particle_clusters}

\end{document}